\documentclass[12pt]{article} 
\usepackage{epsfig}
\usepackage{color}
\usepackage{amsmath}
\usepackage{amssymb}
\usepackage{fullpage}

\begin{document}

%\begin{flushright}
%14.5.2012
%\end{flushright}

\centerline{\bf \Large Scattering phase shifts for  two particles}

\vspace{0.3cm}

 \centerline{\bf \Large of different mass and non-zero total momentum}

\vspace{0.3cm}

 \centerline{\bf \Large in lattice QCD }

\vspace{0.8 cm}

\centerline{ \large Luka Leskovec$^{(a)}$ and Sasa Prelovsek$^{(b,a)}$}

\vspace{0.8cm}

\centerline{\it (a) Jozef Stefan Institute, Jamova 39, 1000 Ljubljana, Slovenia}

\vspace{0.2cm}

\centerline{\it (b) Faculty of Mathematics and Physics, University of Ljubljana, Jadranska 19, Ljubljana, Slovenia}

\vspace{0.3cm}

\centerline{e-mail: {\tt sasa.prelovsek@ijs.si}}

\vspace{1cm}

{\bf Abstract}

\vspace{0.2cm}

We derive  the relation between the scattering phase shift and the two-particle energy in the finite box, which is relevant for extracting the strong phase shifts in lattice QCD. We consider elastic scattering of two particles with different mass and with non-zero total momentum  in the lattice frame.  This is a generalization of the L\"uscher formula, which considers zero total momentum, and the generalization of Rummukainen-Gottlieb's formula, which considers degenerate particles with non-zero total momentum.  We focus on the most relevant  total momenta in practice, i.e. $\mathbf{P}=(2\pi/L)~e_z$ and $\mathbf{P}=(2\pi/L)~(e_x+e_y)$  including their multiples and permutations. We find that the $P$-wave phase shift can be reliably extracted from the two-particle energy if the phase shifts for $l\geq 2$  can be neglected, and we present the corresponding relations. The reliable extraction of S-wave phase shift is much more challenging since $\delta_{l=0}$ is always accompanied by $\delta_{l=1}$ 
in the phase shift relations, and we propose strategies for estimating $\delta_{l=0}$. We also propose the quark-antiquark and meson-meson interpolators that transform according the considered irreducible representations.

\vspace{1cm}

\section{Introduction}

The phase shifts for strong elastic scattering of two hadrons  encode the basic knowledge on the strong interaction between two hadrons, which is non-perturbative in its nature. The phase shift $\delta_l$ is related to the phase between the out-going and the in-going $l$-wave in the region outside the interaction range, and  parametrizes our ignorance of the  complicated form of this interaction. The phase shift  indicates whether the interaction is attractive or repulsive, what is its strength as well as the range, and it provides the value of the scattering length.  
The knowledge of the phase-shift also serves to determine the masses and the width of the resonance that appears in the $l$-wave: $\delta_l=\pi/2$ at the resonance peak, while the sharpness of the rise allows the determination of the resonance width according to Breit-Wigner type functional form. In fact, the only feasible method for determining the resonance width on the  lattice at present goes through the determination of the phase shift. 

\vspace{0.2cm}

Two decades ago, L\"uscher proposed how do determine  phase shifts for elastic scattering in a lattice simulation \cite{luscher_npb}. He derived the  phase shift relation, that relates  the two-particle energy $E$ on a lattice of size $L$ and the infinite volume scattering phase shift  $\delta_l(s)$, where $s=E^2-\mathbf{P}^2$ and $\mathbf{P}$ is the total three-momentum of two particles. If one determines two-particle energy $E$ from a lattice simulation with a given momentum $\mathbf{P}$,  one can extract the phase shift $\delta(s)$ at particular $s=E^2-\mathbf{P}^2$ using this relation. L\"uscher considered only the case $\mathbf{P}=0$. The explicit phase shift relations for higher $l$ at $\mathbf{P}=0$ ware written down in \cite{savage_phase}.

In order to determine $\delta(s)$ at more values of $s=E^2-\mathbf{P}^2$, one better considers also the case $\mathbf{P}\not =0$.  The phase shift relations for this case were derived  in \cite{RG,christ,sharpe,feng_proc}, but these consider the scattering of two particles with equal mass ($m_1=m_2$) \footnote{The determinant condition (50) is derived for general $m_{1,2}$ in \cite{sharpe}, while the function $F$ that enters in it is provided for $m_1=m_2$}. 

In this paper we derive the phase shift relations for the general scattering of two particles with different mass ($m_1\not =m_2$) and with non-zero total momentum ($\mathbf{P}\not =0$). A first step in this direction was made by Davoudi and  Savage \cite{savage} and by Fu \cite{fu}, where the phase shift relation for the irreducible representation $A_1$ was written down: the relation in \cite{savage} takes into account only the $S$-wave interaction and neglects all higher partial waves, while the relation in \cite{fu} takes into account $S$ and $P$ wave.  However, the $A_1$ representation is the least interesting in practice as it mixes $S$ and $P$-wave phase shifts in one relation if $\mathbf{P}\not =0$ and $m_1\not =m_2$, making it difficult to reliably extract any of the two. We derive the phase shift relations also for the other irreducible representations entering in $S$ or $P$-wave scattering with total momentum $\mathbf{P}=(2\pi/L)e_z$ and $\mathbf{P}=(2\pi/L)(e_x+e_y)$: these representations do not mix  $S$ and $P$-wave phase shifts. We also propose the form of lattice interpolators that transform according to these irreducible representations.  

The analogous case of a moving bound state, which is composed of two particles with different mass, has been recently explored in \cite{savage,meissner}. The corresponding finite volume corrections for the S-wave interaction of two particles has been derived in non-relativistic quantum mechanics  \cite{meissner} and in quantum field theory \cite{savage}.

\vspace{0.3cm}

There have been a number of lattice simulations that extracted the phase shift using  $\mathbf{P}=0$, or considering $\mathbf{P}\not = 0$ but $m_1=m_2$.
The $\pi\pi$ scattering with $I=2$ has been studied most frequently; see for example \cite{orginos_pipi_two,ishizuka_pipi_two} for $S$-wave  at $\mathbf{P}\not =0$ and  \cite{dudek_pipi_two} for $D$-wave  at $\mathbf{P}=0$. 
The $\rho$ resonance in $\pi\pi$ scattering with $I=1$ is the only resonance that has been clearly observed in the lattice studies of the phase shifts, which allowed the determination of its mass and width \cite{rho_cppacs,rho_etmc,rho_our,rho_pacscs,rho_qcdsf,rho_bmw}. The preliminary results for the challenging $\pi\pi$ scattering with $I=0$ were presented  in \cite{liu_pipi_zero,fu_pipi_zero}. The $K\pi$ phase shift was extracted from the ground state with $\mathbf{P}=0$ \cite{kpi}, and the results on $I=3/2$ are more reliable than the $I=1/2$ ones. The analytic studies of the phase shift relations with $\mathbf{P}=0$ that may reveal the nature of the scalars mesons in these channels were presented in \cite{rusetsky}.  
The $\rho\pi$ scattering and the related  $a_1$ resonance  ware simulated  in \cite{a1_our},  while the corresponding phase shift-relations for the scattering of unstable particles was analytically studied in \cite{a1_oset}. The preliminary results for channels including charmed and charmonium states were presented in \cite{liuming}, while $D^*D_1$ scattering relevant for $Z^+(4430)$ was simulated in \cite{Zplus}. Recent review of  the applications, including also baryons, multi-particle interactions and bound states was presented in \cite{orginos_hadron_int}. 

There are however many interesting channels, where two scattering particles have different mass, and the simulations at non-zero total momentum would provide the valuable information on the corresponding phase shifts. To our knowledge, the phase shifts have not been extracted from lattice in such a case, and we provide analytical tools that would enable that in the near future. 

\vspace{0.2cm}

In Section 2 we first consider two non-interacting particles in the finite volume, then we consider the interacting particles and write down the general phase shift relation. In  Section 3 we simplify the general phase shift relations by considering the discrete symmetries. First we focus on the case of total momentum $\mathbf{P}=(2\pi/L)(e_x+e_y)$, write down the phase shift relations for three irreducible representations that appear in $S$ or $P$-wave scattering, discuss the strategies for extracting the phase shifts $\delta_{l=0,1}$ and provide the quark-antiquark and meson-meson interpolators that transform according to these representations. Then we repeat the same steps for the total momentum $\mathbf{P}=(2\pi/L)e_z$. We end with conclusions. The Appendix provides the derivation of the expression for the generalized zeta function $Z_{lm}^{\mathbf{d}}(1;q^2)$ for $m_1\not =m_2$, that is appropriate for  numerical evaluation.

\section{Two particles in a finite volume}

We consider a square lattice box of volume $L^3$ with periodic boundary conditions in all three spatial directions, while time extent is  infinite. We assume continuous space-time and we do not consider discretization errors due to the finite lattice spacing $a$ in actual simulations with a given action. There are two particles with total three-momentum $\mathbf{P}$ in such a box, and the total momentum has to satisfy the periodic boundary condition
\begin{equation}
\label{d}
\mathbf{P}=\mathbf{p}_1+\mathbf{p}_2\equiv \frac{2\pi}{L}\mathbf{d}\ , \qquad \mathbf{d}\in Z^3~.
\end{equation} 
The main task is to derive  the total energy  $E$ of the these two particles, where $E$ refers to the energy measured by the observer that is at rest with  respect to the lattice frame (LF), i.e. lattice square box. 
First we consider the non-interacting case, which is trivial. Then we turn to the interacting case, where the energy $E$  depends on the scattering phase shifts $\delta_l$ in the $l$-th  partial wave. This relation will ultimately allow for the determination of $\delta_l$ from the energies determined by lattice simulations in a finite box. 

The scattering in partial wave $l$ refers to the center-of-momentum  frame (CMF), which moves with the velocity
\begin{equation} 
\mathbf{v}=\frac{\mathbf{P}}{E}\quad,\qquad \gamma=\frac{1}{\sqrt{1-\mathbf{v}^2}}
\end{equation}
with respect to the lattice frame. Therefore we need to consider the physical system in CMF, where the quantities  will be denoted by $*$. The Lorentz transformation between two systems is performed by $\hat \gamma$, which acts on a general vector $\mathbf{u}$ as 
\begin{equation}
\label{hat_gam}
\hat \gamma \mathbf{u}=\gamma \mathbf{u}_\parallel + \mathbf{u}_\perp ~,\quad \hat \gamma^{-1} \mathbf{u}=\gamma^{-1} \mathbf{u}_\parallel + \mathbf{u}_\perp ~,\qquad \mathbf{u}_\parallel =\tfrac{\mathbf{u}\cdot \mathbf{v}}{|\mathbf{v}^2|}~\mathbf{v}~,\quad \mathbf{u}_\perp=\mathbf{u}-\mathbf{u}_\parallel
\end{equation}
so it  preserves the component perpendicular to $\mathbf{v}$ and modifies the component parallel to $\mathbf{v}$. The lattice square box is deformed to some general parallelepiped, and its shape depends on the direction of $\mathbf{P}$. The  
two-particle wave functions in CMF will ``see'' the lattice box in the shape of this parallelepiped and the technical difficulty is that the periodic boundary condition on the CMF wave functions has to be enforced with respect to this parallelepiped. 

\subsection{Non-interacting case}

In the non-interacting case there are two major simplifications: the momenta of the individual particles also satisfy the periodic boundary condition and  the energy is the sum of the individual energies 
\begin{equation}
\label{E_noninteract}
E=\sqrt{\mathbf{p}_1^2+m_1^2}~+\sqrt{\mathbf{p}_2^2+m_2^2}\ ,\qquad  \mathbf{p}_1+\mathbf{p}_2=\mathbf{P}~,\qquad \mathbf{p}_1=\frac{2\pi}{L}\mathbf{n} ~,\quad \mathbf{p}_2=\frac{2\pi}{L} \mathbf{n'}~,\qquad \mathbf{n},\mathbf{n'}\in Z^3~.
\end{equation}
This already provides the two-particle discrete energy spectrum in absence of interactions.

\begin{figure}[bt]
\begin{center}
\includegraphics*[width=0.7\textwidth,clip]{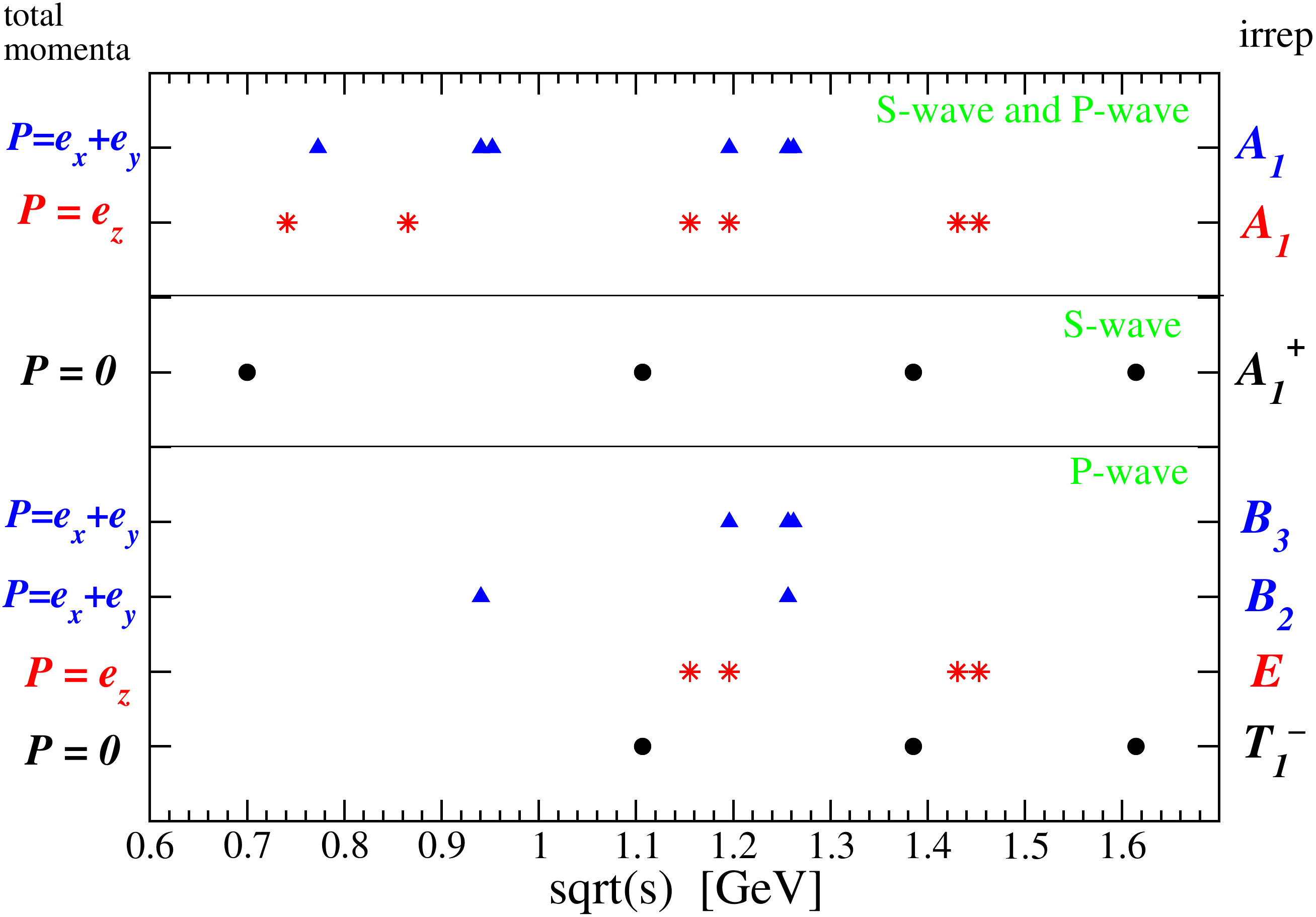} 
\end{center}
\caption{Values of the allowed $\sqrt{s}=\sqrt{E^2-P^2}$ for non-interacting scattering states of two particles with mass $m_1=200$ MeV, $m_2=500$ MeV and total momentum $\mathbf{P}$   in the lattice box of $L=3$ fm. Simulations at $P=0$ provide only the values of $\sqrt{s}$ given by the circles, while the simulations with $\mathbf{P}=\tfrac{2\pi}{L}e_z$ and $\mathbf{P}=\tfrac{2\pi}{L}(e_x+e_y)$ provide also the values of $\sqrt{s}$ given by the stars and triangles, respectively. Each line corresponds to a definite irreducible representation in CMF, and corresponding $\sqrt{s}$ are obtained from $E$ (\ref{E_noninteract}) and choices of $\mathbf{p}_{1,2}$ in  (\ref{interpolators_c2v}) and (\ref{interpolators_c4v}).  }\label{fig_example}
\end{figure}

The energies of the interacting scattering states will be slightly shifted with respect to the non-interacting case (\ref{E_noninteract}). However the non-interacting case already gives us a rough estimate of the expected spectrum of scattering states  and the corresponding  values of $s=E^2-P^2$ in a simulation with given total momentum $\mathbf{P}$. The approximate knowledge on the allowed values of $s$ is very valuable, since such simulation would provide values of  phase shifts $\delta(s)$ at those values of $s$. 
The allowed values  of $\sqrt{s}$ for the non-interacting scattering states with $p_1\leq \sqrt{3}~ \tfrac{2\pi}{L}$ and $p_2\leq \sqrt{3}~ \tfrac{2\pi}{L}$  are presented in Fig. \ref{fig_example}. In this example  we take $m_1=200$ MeV, $m_2=500$ MeV (possible values of $m_\pi$ and $m_K$ in the present lattice simulations) and $L=3$ fm. The simulations with $\mathbf{P}=0$ will provide only the values of S-wave and P-wave phase shifts $\delta(s)$ at $\sqrt{s}\simeq \sqrt{m_1^2+n(\tfrac{2\pi}{L})^2}+ \sqrt{m_2^2+n(\tfrac{2\pi}{L})^2}$ shown by circles; note that the lowest scattering state $P_1(0)P_2(0)$ is not present in P-wave. Simulations at $\mathbf{P}=\tfrac{2\pi}{L}e_z$ and $\mathbf{P}=\tfrac{2\pi}{L}(e_x+e_y)$ will provide the values of S-wave\footnote{It will be shown in the following sections that the S-wave appears only in the irreducible representation $A_1$ for $\mathbf{P}\propto e_z$ and $\mathbf{P}\propto e_x+e_y$. P-wave will also appear in this irrep, so extracting S-wave phase shift is challenging, as discussed in section \ref{symmetries_c2v}. } and P-wave phase shifts at additional values of $\sqrt{s}$ given by the stars and triangles, respectively. Those values of $\sqrt{s}=\sqrt{E^2-\mathbf{P}^2}$ are obtained simply by using the energies $E$ (\ref{E_noninteract}) for certain values of $\mathbf{p}_{1}\in \tfrac{2\pi}{L}\mathbf{n}$ and $\mathbf{p}_{2}\in \tfrac{2\pi}{L}\mathbf{n'}$.  The allowed combinations of $\mathbf{p}_1$ and $\mathbf{p}_2$ will be understood only after we consider the symmetries of the two-particle system in CMF. For each irreducible representation they can be read off from the $P_1P_2$ interpolators given in (\ref{interpolators_c2v}) and (\ref{interpolators_c4v}).  

\vspace{0.2cm}

Since symmetries in CMF frame will be important, 
 we also need the values of the allowed momenta $\mathbf{p}^*$ in the CMF frame for non-interacting case
\begin{equation}
\mathbf{p}^*=\mathbf{p}_1^*=-\mathbf{p}_2^*\qquad p^*=|\mathbf{p}^*|
\end{equation}
to study the scattering. We extract $\mathbf{p}^*$  from $\mathbf{p}_1=\tfrac{2\pi}{L}\mathbf{n}$ using
\begin{align}
\label{pstar}
\mathbf{p}_1&=\hat \gamma (\mathbf{p}^*+\mathbf{v}E_1^*)\ ,\qquad \mathbf{p}_2=\hat \gamma (-\mathbf{p}^*+\mathbf{v}E_2^*) \qquad \mathrm{so}\nonumber\\
 \mathbf{p}^*&=\hat \gamma^{-1} \mathbf{p}_1-\mathbf{v}E_1^*=\hat \gamma^{-1} [\mathbf{p}_1-\gamma \mathbf{v}E_1^*] =\hat \gamma^{-1} \biggl(\mathbf{p}_1-\gamma ~\frac{2\pi \mathbf{d}}{L E} \frac{E^*}{2}~\bigl[1+\frac{m_1^2-m_2^2}{E^{*2}}\bigr]\biggr)\nonumber\\
\mathbf{p}^*&=\hat \gamma^{-1} ~(\mathbf{p}_1-\tfrac{1}{2}~A ~\mathbf{P})\ ,
\end{align}
where in the second step $E_1=(E^*/2)[1+(m_1^2-m_2^2)/E^{*2}]$ is expressed in terms of energy in CMF
\begin{equation}
E^*=\sqrt{p^{*2}+m_1^2}+\sqrt{p^{*2}+m_2^2}=\gamma^{-1}~E~
\end{equation}
and we have defined coefficient $A$ 
\begin{equation}
\label{A}
A\equiv 1+\frac{m_1^2-m_2^2}{E^{*2}}
\end{equation} 
which is different from $1$ only when $m_1\not = m_2$. 
We express the values of $\mathbf{p^*}$ in terms of the dimensionless CMF momentum $\mathbf{q}$  
\begin{equation}
\label{q}
\mathbf{p}^*\equiv \frac{2\pi}{L}\mathbf{q}
\end{equation}
and the allowed values of $\mathbf{q}$ in the non-interacting case are 
\begin{equation}
\label{q_equal_r}
\mathbf{q}=\mathbf{r} \qquad \mathbf{r}\in P_d\qquad \mathrm{(for \ non-interacting\ case)}
\end{equation}
where the $\mathbf{r}\in P_d$ is set of vectors given by the mesh obtained combining  (\ref{pstar},\ref{q},\ref{q_equal_r}) and $\mathbf{p}_1=\tfrac{2\pi}{L}\mathbf{n}$ 
\begin{equation}
\label{P_d}
P_d= \{~\mathbf{r}\quad | \quad \mathbf{r}=\hat{\gamma}^{-1}(\mathbf{n}-\tfrac{1}{2}~A~\mathbf{d})~\}\ , \qquad \mathbf{n}\in Z^3~
\end{equation}
which agrees with \cite{savage,fu}\footnote{We have a different sign than \cite{fu} in $P_d$, but both signs lead to the same (infinite) mesh of points.}.   
The symmetries under which this set of points is invariant will play a major role later on. 
The equality $\mathbf{q}=\mathbf{r}$ (\ref{q_equal_r}) will be modified by the two-particle interactions in the finite volume.

\subsection{Interacting case}

Now we consider the elastic scattering of two interacting particles with spin 0 in a finite box using relativistic quantum mechanics, along the lines of Rummukainen-Gottlieb  that consider $m_1=m_2$ \cite{RG}, and 
Fu that presented the analogous derivation for $m_1\not = m_2$ \cite{fu}. 

For the case $m_1=m_2$, the quantum mechanics result of \cite{RG} was subsequently reproduced   using the Bethe-Salpeter equation \cite{christ} and using the quantum field theory \cite{sharpe}.  The phase shift relations from all approaches agree when 
 one neglects the terms, that are exponentially suppressed with the box size $L$ in the quantum field theory. The phase shift relations derived here can therefore be applied if $L$  in the simulation is large enough that the terms of the order of  $e^{-m_\pi L}$ can be neglected.

We need to find the two-particle energies $E$ in the finite box in the presence of the 
 potential $V(\mathbf{x}^*)$, which depends on their relative distance $\mathbf{x}^*= \mathbf{x}_1^*-\mathbf{x}_2^*$. The strong potential between 
two hadrons in not known ab-initio in QCD, so one can not analytically calculate the eigen-energies $E$ which satisfy $\hat H \psi(x_1,x_2)=E\psi(x_1,x_2)$, but rather determines eigen-energies $E$ in lattice QCD, which incorporates fundamental QCD interactions. 

However, one can analytically consider the two-particle wave functions in the {\it exterior region}, where the potential drops to zero
\begin{equation}
V(\mathbf{x}^*)=0\qquad \mathrm{for}\qquad |\mathbf{x}^*|=x^*>R
\end{equation}
and  we assume the interaction is of finite range\footnote{Presence of the exterior region is not necessary in the  quantum field derivation  \cite{sharpe}.} $R<L/2$. In the exterior region,  
the two-particle wave function will satisfy $\hat H_{free} \psi_{free}(x_1,x_2)=E\psi_{free}(x_1,x_2)$ with the same eigen-energy $E$ as in the case of Hamiltonian  $\hat H$ with interactions. The relation $\hat H_{free} \psi_{free}(x_1,x_2)=E\psi_{free}(x_1,x_2)$ in the exterior region of  CMF has a form of the well-known Helmholtz equation 
\begin{align}
&(\nabla^2 + p^{*2})\phi_{CM}(\mathbf{x}^*)=0 \qquad x^*>R\label{helmholtz1}\\
& ~\nonumber\\
 &E^*=\sqrt{p^{*2}+m_1^2}+\sqrt{p^{*2}+m_2^2}=\gamma^{-1}~E~\label{helmholtz2}
\end{align}  
and the total energy $E^*$ is just a sum of both individual energies in this region. 

The only effect of the interior region $x^*<R$ on the 
free solutions in the exterior region is that $\phi_{CM}(\mathbf{x}^*)$ will depend on the phase shifts $\delta_l(p^*)$, which are related to the phase between the out-going and in-going $l$-wave and  parametrize our ignorance of the exact form of the potential $V(\mathbf{x}^*)$. A free solution  in the exterior region with  momentum $p^*$, which is  shifted by phase shift $\delta_l(p^*)$, will satisfy the periodic boundary condition only for some specific values of $p^*$, which fulfill certain relation between $p^*$, $\delta_l(p^*)$ and  $L$. This relation is the analog of the L\"uscher formula we are looking for: 
it will provide $\delta_l(p^*)$ if one determines the momentum $p^*$ in the exterior region on a lattice of size $L$. This exterior momentum $p^*$ is extracted via relation (\ref{helmholtz2})
%$E^*=\sqrt{p^{*2}+m_1^2}+\sqrt{p^{*2}+m_2^2}=\gamma^{-1}~E$
from the energy $E$ of two strongly interacting particles, where $E$ is directly measured  with a lattice QCD simulation in a box of size $L$.\\

Before imposing the boundary conditions in the finite volume, let us review the 
familiar solutions of the Helmholtz equation for given $p^*$ in the infinite volume
\begin{equation}
\label{sol_infinite_vol}
\phi_{CM}(\mathbf{x}^*)=\sum_{l,m} c_{lm}~Y_{lm}(\theta,\varphi)~[a_l(p^*)j_l(p^*x^*)+b_l(p^*)n_l(p^*x^*)]\ , \qquad x^*>R~,
\end{equation}
which apply in the exterior region. 
The phase shift $\delta_l(p^*)$ in the continuum is commonly defined through the  ratio of the out-going\footnote{We apply definition of $n_l(x)\stackrel{x\to \infty}{\longrightarrow} \cos (x-l\pi/2)/x$ which agrees with \cite{messiah} and \cite{luscher_npb}, but differs in sign with 
commonly used definitions.}
 $l$-wave $j_l-i n_l$ and the in-going wave $j_l+i n_l$ with momentum $p^*$
\begin{equation}
\label{phase_def}
e^{2i\delta_l(p^*)}\equiv \frac{a_l(p^*)+ib_l(p^*)}{a_l(p^*)-ib_l(p^*)}~.
\end{equation}
We will use the same definition of the phase shift in the finite volume, but there the wave function will not be so simply expressed in terms of $j_l$ and $n_l$ due to the boundary conditions. \\

Now we turn to the solutions $\phi_{CM}(\mathbf{x}^*)$ in the exterior region, that satisfy the Helmholtz equation  and also the boundary condition at finite $L$. We consider the case of the periodic boundary condition in the lattice frame 
\begin{equation}
\psi(\mathbf{x}_1,\mathbf{x}_2)=\psi(\mathbf{x}_1+\mathbf{n}_1L~,~\mathbf{x}_2+\mathbf{n}_2L)
\end{equation}
which is most commonly used in the actual simulations. These   boundary conditions impose that $\phi_{CM}(\mathbf{x}^*)$  need to satisfy the so-called $\mathbf{d}$-periodic boundary condition, which was derived by Fu \cite{fu} for $\mathbf{P}\not =0$ and $m_1\not =m_2$ using the Lorentz transformation between the two frames\footnote{Fu \cite{fu} has a different sign here, but this represents exactly the same boundary condition since 
$(-1)^{A\mathbf{n}\cdot \mathbf{d}}=\exp(\pm i\pi \mathbf{n}\cdot \mathbf{d})\exp(\pm i\pi \tfrac{m_1^2-m_2^2}{E^{*2}}\mathbf{n}\cdot \mathbf{d})$ where each of two signs can be $+$ or $-$. We choose different sign than Fu as it is more in line with our definition of $P_d$ (\ref{P_d}). }:
\begin{equation}
\label{bc}
\phi_{CM}(\mathbf{x}^*)=(-1)^{A\mathbf{n}\cdot \mathbf{d}}~\phi_{CM}(\mathbf{x}^*+\hat \gamma \mathbf{n} L)\qquad \mathbf{n}\in Z^3\ .
\end{equation}

A simple example, that satisfies the Helmholtz equation and the $\mathbf{d}$-periodic boundary condition, is the Green function 
\begin{equation}
\label{sol1}
G^{\mathbf{d}}(\mathbf{x}^*,p^{*2})=\gamma^{-1}L^{-3}\sum_{k=\tfrac{2\pi}{L}\mathbf{r},~\mathbf{r}\in P_d} \frac{e^{i\mathbf{k}\cdot \mathbf{x}^*}}{\mathbf{k}^2-p^{*2}}
\end{equation}
where $P_d$ is a mesh of points $P_d$ (\ref{P_d}).  
 Other solutions that satisfy the Helmholtz equation and the boundary conditions (\ref{bc}) are \cite{luscher_npb,RG,fu} \footnote{For our purpose ${\cal Y}_{lm}(\nabla)$  was most conveniently applied if both $G^{\mathbf{d}}(\mathbf{x}^*,p^{*2})$ and  ${\cal Y}_{lm}(\tfrac{\partial}{\partial x^*},\tfrac{\partial}{\partial y^*},\tfrac{\partial}{\partial z^*})$ are expressed in terms of the cartesian coordinates.}
\begin{equation}
\label{sol2}
G_{lm}^{\mathbf{d}}(\mathbf{x}^*,p^{*2})={\cal Y}_{lm}(\nabla)~G^{\mathbf{d}}(\mathbf{x}^*,p^{*2})\ ,\qquad {\cal Y}_{lm}(\mathbf{x})\equiv x^l~Y_{lm}(\theta,\varphi)~. 
\end{equation}
The solutions $G_{lm}^{\mathbf{d}}$ form a complete basis, as shown by L\"uscher for $\mathbf{d}=0$ \cite{luscher_npb}, and the general solution $\phi_{CM}(\mathbf{x}^*)$ can be expanded in terms of them 
\begin{equation}
\label{sol3}
 \phi_{CM}(\mathbf{x}^*)=\sum_{l,m}v_{lm}~G_{lm}^{\mathbf{d}}(\mathbf{x}^*,p^{*2}) ~\ , \qquad x^*>R\ .
\end{equation}

The technical difficulty arises from the fact that the solutions (\ref{sol3}) satisfy boundary conditions (\ref{bc}) related to the parallelepiped in CMF, while the phase shifts $\delta_l$ are related to the coefficients in front of 
the out-going  $j_l-i n_l$ and in-going $j_l+in_l$ spherical waves as in the infinite volume (\ref{phase_def}). So one needs  to express the $\mathbf{d}$-periodic solutions (\ref{sol2}) in terms of the spherical Bessel functions $j_l$ and $n_l$. That can be fortunately done in analogous way 
as performed for $\mathbf{d}=0$ by L\"uscher \cite{luscher_npb} and we omit the derivation here
\begin{equation}
\label{sol4}
G_{lm}^{\mathbf{d}}(\mathbf{x}^*,p^{*2})=\frac{(-1)^l (p^*)^{l+1}}{4\pi}\biggl[n_l(p^*x^*)~Y_{lm}(\theta,\varphi)~+~\sum_{l'=0}^{\infty}\sum_{m'=-l'}^{l'}~{\cal M}_{lm,l'm'}^{\mathbf{d}}(q^2)~j_{l'}(p^*x^*)~Y_{l'm'}(\theta,\varphi)\biggr]
\end{equation}
where ${\cal M}_{lm,l'm'}^d(q^2)$ are calculable matrices for given $l,~m,~l',~m',~\mathbf{d},~q\!=\!Lp^*/(2\pi)$ and $A$ (\ref{A}) and the explicit expression will be given in the next section. The $\theta$ and $\varphi$ are polar angles of $\mathbf{x}^*$. 

The general solution in the exterior region $\phi_{CM}(\mathbf{x}^*)$ (\ref{sol3}) is obtained by inserting (\ref{sol4}), and one needs to relate this to the form (\ref{sol_infinite_vol}) in order to extract the phase shifts defined by (\ref{phase_def})  
\begin{align}
\label{sol5}
\phi_{CM}(\mathbf{x}^*)&=\sum_{lm}v_{lm} \frac{(-1)^l (p^*)^{l+1}}{4\pi}\biggl[n_l(p^*x^*)~Y_{lm}(\theta,\varphi)~+~\sum_{l',m'}~{\cal M}_{lm,l'm'}^{\mathbf{d}}(q^2)~j_{l'}(p^*x^*)~Y_{l'm'}(\theta,\varphi)\biggr]\nonumber\\
&=\sum_{l,m} c_{lm}~Y_{lm}(\theta,\varphi)~[a_l(p^*)j_l(p^*x^*)+b_l(p^*)n_l(p^*x^*)]\ ,\qquad x^*>R\ .
\end{align}
By equating the terms in from of $Y_{lm}n_l$ and $Y_{lm}j_l$ we get two  relations
\begin{equation}
v_{lm}\frac{(-1)^l (p^*)^{l+1}}{4\pi}=c_{lm} ~b_l(p^*)\ ,\qquad  \sum_{l',m'} v_{l'm'}\frac{(-1)^{l'} (p^*)^{l'+1}}{4\pi}{\cal M}_{l'm',lm}(q^2)=c_{lm} ~a_l(p^*)~
\end{equation}
and $v_{lm}$ can be expressed from the first relation and inserted into the second
\begin{equation}
\sum_{l',m'} c_{l'm'} \bigl[b_{l'}(p^*)~{\cal M}_{l'm',lm}(q^2) - a_{l'}(p^*)\delta_{ll'}\delta_{mm'}\bigr]=0 \ .
\end{equation}
This linear system has nontrivial solution for $c_{l'm'}$ only if 
\begin{equation}
\label{sol6}
det(BM-A)=0\qquad A_{lm,l'm'}\equiv a_l(p^*)\delta_{ll'}\delta_{mm'}\ , \quad B_{lm,l'm'}\equiv b_l(p^*)\delta_{ll'}\delta_{mm'}\ 
\end{equation}
where $A$ and $B$ are defined as diagonal matrices related to coefficients $a_l$ and $b_l$ \cite{luscher_npb} and they finally provide the information on the phase $\delta_l$ defined by (\ref{phase_def})
\begin{equation}
e^{2i\delta}=\frac{A+iB}{A-iB}~.
\end{equation}
By dividing (\ref{sol6}) by $det(A-iB)$, which is non-zero \cite{luscher_npb}, one obtains the final relation between the diagonal matrix $e^{2i\delta}$ and (in general) non-diagonal matrix  $M$ 
   \begin{align}
\label{det} 
&det[e^{2i\delta}(M-i)-(M+i)]=0\ , \qquad \qquad \quad l,l'\leq l_{max} \\
~\nonumber\\
\quad M_{lm,l'm'}\equiv & {\cal M}^{\mathbf{d}}_{lm,l'm'}(q^2)\ ,\quad [e^{2i\delta}]_{lm,l'm'}\equiv e^{2i\delta_l(p^*)}\delta_{ll'}\delta_{mm'}\ .\nonumber 
\end{align}
 This  condition is the heart of the phase shift relation  and relates the energy $E$ (or $q$) measured on the lattice to the unknown phases $\delta_l(p^*)$ via the calculable matrix elements ${\cal M}^{\mathbf{d}}_{lm,l'm'}(q^2)$, that will be given in the next subsection. The energy level $E$ will provide the information on the phase shift $\delta_l(p^*)$ at CMF momentum $p^*$, that is related to $E$ via (\ref{helmholtz2}). 

If $\delta_l=0$ for $l> l_{max}$, the relation (\ref{det}) needs to be satisfied for the truncated square matrices with $l,l'\leq l_{max}$, as shown in \cite{sharpe}. 

The determinant of the block-diagonal matrix is a product of determinants for separate blocks. So the determinant condition will get simplified when $M$ will be  written in such basis that leads to a block-diagonal form of $M$ and therefore block-diagonal form of $e^{2i\delta}(M\!-\!i)-(M\!+\!i)$.  

\subsection{ Definitions of ${\cal M}^{\mathbf{d}}_{lm,l'm'}$ and $Z_{lm}^d$ for $m_1\not = m_2$}

Finally we write down the explicit expression for ${\cal M}^{\mathbf{d}}_{lm,l'm'}(q^2)$, that were introduced while expanding $G^{\mathbf{d}}_{lm}$ in terms of $j_l$ and $n_l$ (\ref{sol4}) \cite{luscher_npb,RG,fu}
\begin{align}
\label{M_def}
{\cal M}^{\mathbf{d}}_{lm,l'm'}(q^2)&\equiv \frac{(-1)^l}{\gamma\pi^{3/2}}\sum_{j=|l-l'|}^{l+l'}\sum_{s=-j}^j \frac{i^j}{q^{j+1}}Z_{js}^{\mathbf{d}}(1;q^2)~C_{lm,js,l'm'} \ ,\\
C_{lm,js,l'm'}&\equiv (-1)^{m'}i^{l-j+l'}\sqrt{(2l+1)(2j+1)(2l'+1)}\biggl({l\atop m}{j\atop s}{l'\atop -m'}\biggr) \biggl({l\atop 0}{j\atop 0}{l'\atop 0}\biggr) \ ,\nonumber
\end{align}
where $C_{lm,js,l'm'}$ is expressed in terms of the $3j$-Wigner symbols. The modified zeta function 
is defined as in \cite{savage,fu}
\begin{equation}
\label{Z_def}
Z^{\mathbf{d}}_{lm}(s;q^2)\equiv \sum_{\mathbf{r}\in P_d} \frac{{\cal Y}_{lm}(\mathbf{r})}{(\mathbf{r}^2-q^2)^s} \ ,
\end{equation}
where $P_d$ is the mesh of points defined in (\ref{P_d}), and ${\cal Y}_{lm}$ is defined in (\ref{sol2}).  In the special case $\mathbf{d}=0$, the definition of  the zeta function (\ref{Z_def}) agrees with the original definition by L\" uscher \cite{luscher_npb}. 
The zeta function depends on $l,~m$, $q^2=(Lp^*/2\pi)^2$, $\mathbf{d}$  and $A$ (\ref{A}). The $Z_{00}$ is finite only for $s>3/2$, but the divergence is not physical as it cancels in the difference between the finite and infinite volume result, as explained in the Appendix A, where $Z_{00}^{\mathbf{d}}$ is obtained by analytical continuation from $s>3/2$ to $s=1$. The $Z_{lm}^{\mathbf{d}}(1;q^2)$ is finite for $l\not =0$, but the sum (\ref{Z_def}) converges to slowly for practical evaluation.  In Appendix A we derive an expression that is suitable for the numerical evaluation and reproduces the known result in the special case $m_1=m_2$ \cite{RG,feng_proc,rho_our}.  

\subsection{General form of $M$ for $l_{max}=1$}

In this paper we are especially interested in extracting the $S$ and $P$-wave phase shifts $\delta_{l=0,1}(p^*)$ 
for the scattering of two particles with different mass. This problem is significantly simplified 
if the scattering phases for the partial waves with $l>l_{max}=1$ are small and can be neglected, i.e. 
 we will assume that $\delta_{l>1}=0$. This is generally true for small $p^*$, where higher partial waves are generally suppressed by $\delta(p^*)\propto (p^{*})^{2l+1}$   
and is often true also for a range of $p^*$ if there is no $D$-wave resonance in the vicinity. The phases for the higher partial waves were explicitly found to be small for $p^*\geq 1$ GeV in the simulation \cite{dudek_pipi_two} of the non-resonant channel  $\pi\pi$ with $I=2$. 

Assuming  $\delta_{l>1}=0$, the matrix $M$ is $4\times 4$ matrix in the basis $lm=00,10,11,1-1$ and the expression  ${\cal M}_{lm,l'm'}^{\mathbf{d}}$ (\ref{M_def})  leads to the following form for general $\mathbf{d}$ 
\begin{equation}
\label{M_gen}
M={\cal M}_{lm,l'm'}^{\mathbf{d}} = \bordermatrix{~ & 00        & 10                    & 11                       & 1-1\cr
                            00 & w_{00} & i\sqrt{3}w_{10} & i \sqrt{3} w_{11} & i\sqrt{3} w_{1-1} \cr
                            10 & -i\sqrt{3}w_{10} & w_{00}+2w_{20} & \sqrt{3} w_{21} & \sqrt{3}w_{2-1} \cr
                            11 & i \sqrt{3}w_{1-1} & -\sqrt{3}w_{2-1} & w_{00}-w_{20} & -\sqrt{6} w_{2-2} \cr
                            1-1 & i \sqrt{3} w_{11} & -\sqrt{3}w_{21} & -\sqrt{6}w_{22} & w_{00}-w_{20} \cr}\ ,
\end{equation}
where we defined $w_{lm}$ to simplify the notation
\begin{equation}
\label{w}
w_{lm}\equiv \frac{1}{\pi^{3/2} \sqrt{2l+1} ~\gamma~ q^{l+1}}~ Z_{lm}^{\mathbf{d}}(1;q^2)\ .
\end{equation}

\section{Consequences of the discrete symmetries of mesh $P_d$}

Some of the matrix elements ${\cal M}_{lm,l'm'}^{\mathbf{d}}$ (\ref{M_gen}) are zero for particular choices of $\mathbf{d}$, which becomes apparent after explicit numerical evaluation of the $w_{lm}\propto Z^{\mathbf{d}}_{lm}(1;q^2)$ given in Appendix A. This is a consequence of the discrete symmetries of the  mesh $P_d$ (\ref{P_d}) in CMF. It is helpful to first study these symmetries and determine the texture of the matrix  ${\cal M}$ (\ref{M_gen}) for particular $\mathbf{d}$ before inserting $M$ to the master determinant condition (\ref{det}). 

So we explore in  this section the group $G$ of the symmetry elements $\hat R$ that leaves the mesh $P_d$ (\ref{P_d}) invariant. Later on we study  the consequences of these symmetries for two most useful types of non-zero momentum $\mathbf{d}\propto e_z$ and $\mathbf{d} \propto e_x+e_y$, or permutations. There are several reasons why symmetry consideration will be helpful:
\begin{itemize}
\item  It will indicate which $Z_{lm}^{\mathbf{d}}$ are zero, purely real or purely imaginary, as indicated above. 
\item The Helmholtz solution $G^{\mathbf{d}}$ (\ref{sol1}) is invariant under the transformations  $\hat R\in G$ due to the sum over $\mathbf{r}\in P_d$. The other solutions $G_{lm}^{\mathbf{d}}$ 
are generated by applying ${\cal Y}_{lm}(\nabla)$, so they transform like $Y_{lm}$   \cite{luscher_npb,feng_proc}
\begin{equation}
\label{G_tr}
G^{\mathbf{d}}_{lm}(\hat R \mathbf{x}^*,p^{*2})=\sum_{m'=-l}^l D_{mm'}^{(l)}(\hat R) ~ G^{\mathbf{d}}_{lm'}(\mathbf{x}^*,p^{*2})\ , 
\end{equation}
where $D_{mm'}^{(l)}(\hat R)$ is defined as a representation in the bases $Y_{lm}$ 
\begin{equation}
\label{Y_tr}
Y_{lm}(\hat R\mathbf{x})=\sum_{m'=-l}^l D_{mm'}^{(l)}(\hat R) Y_{lm'}(\mathbf{x}) 
\end{equation}
and is in general reducible. So  the solutions $G^{\mathbf{d}}_{lm}$ form a representation of group $G$, which is in general reducible.  The energy eigenstates in (\ref{sol3}) are certain linear combinations  of $G^{\mathbf{d}}_{lm}$ that transform according to the irreducible representation of the  $G$;  the representation  $D_{mm'}^{(l)}(\hat R)$  of transformation $\hat R$ (\ref{G_tr}) has the  irreducible block-diagonal form in this basis.  
\item The same linear combinations of $Y_{lm}$ will also lead to the block-diagonal form of $M$, as shown by L\"uscher (see section n5.3 of \cite{luscher_npb}). We will therefore write down $M$ and search for the basis that 
leads to the block-diagonal form\footnote{For higher $l_{max}$ it is probably easier to first determine the basis that makes representation $D$ (\ref{Y_tr}) block diagonal.}. It will turn out that the resulting basis indeed corresponds to the irreducible representations of $D$ (\ref{G_tr},\ref{Y_tr}). 
\item
The determinant condition (\ref{det}) is greatly simplified in the bases where 
$M$ is block-diagonal since the determinant of the block-diagonal matrix is a product of determinants for separate blocks\footnote{Note that $e^{2i\delta}(M\!-\!i)-(M\!+\!i)$ is block diagonal when $M$ is block diagonal, since $e^{2i\delta}$ is diagonal by construction (\ref{det}).}. In this case the determinant condition is simplified to analogous conditions for separate blocks (i.e. irreducible representations). 
\item
The lattice interpolators, that are written down in the lattice frame, have to transform according to the irreducible representation of the group $G$ after transformed to the CMF frame. We will provide 
useful examples of quark-antiquark and meson-meson interpolators, that satisfy this property and be used in the actual simulations to extract the phase shifts.
\end{itemize}
All these reasons prompt us to consider the symmetries of the mesh $P_d$ (\ref{P_d}) for separate cases of $\mathbf{d}$. 

\begin{figure}[bt]
\begin{center}
\includegraphics*[width=0.2\textwidth,clip]{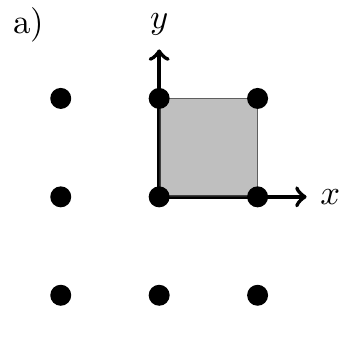} $\quad$
\includegraphics*[width=0.2\textwidth,clip]{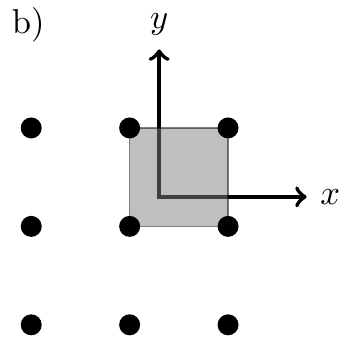} $\quad$
\includegraphics*[width=0.2\textwidth,clip]{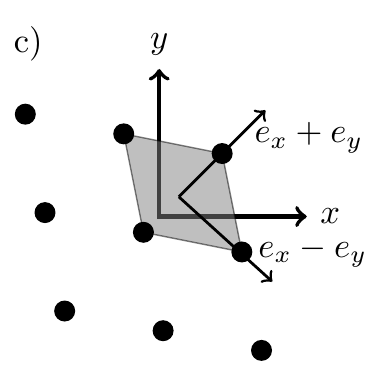}
\end{center}
\caption{The mesh $P_d$ (\ref{P_d}) for $\mathbf{d}=e_x+e_y$ and $m_1\not =m_2$ ($A\not =1$) is plotted in (c), while (a) and (b) show the steps how to get it.  }\label{fig_mesh_110}
\end{figure}

\subsection{$\mathbf{P}= (2\pi/L) (e_x+e_y)$ and consequences of $C_{2v}$ symmetries}
     
     We first explain the case of the momentum $\mathbf{d}=e_x+e_y$ in detail, since it is general enough 
     to illustrate the procedure and since the corresponding group has only few elements. 
     All the results can be easily generalized 
     to the case $\mathbf{d}=N(e_x+e_y)$ or any permutation in the direction. 
     
     We first need to determine the  symmetry transformations $\hat R$ that leave $P_d$ invariant. 
   The mesh $P_d$ (\ref{P_d}) can be visualized in Fig. \ref{fig_mesh_110}c and is obtained in two steps:
   \begin{enumerate}
   \item
    First the cubic mesh  $\mathbf{r}=\mathbf{n}\in Z^3$ in Fig. \ref{fig_mesh_110}a is  shifted by $-\tfrac{1}{2}A \mathbf{d}\propto e_x+e_y$ and since $A\not =1$ for $m_1\not = m_2$ (\ref{A}), the origin is not in the center of the unit cell in $xy$ plane (Fig. \ref{fig_mesh_110}b). The inversion $I$ with respect to the origin is lost as a symmetry at this stage, so the corresponding group $G$ will not contain $I$; 
   this is a major difference with respect to degenerate case $m_1=m_2$, when the origin is at the center of the unit cell in $xy$ plane and $I$ is the symmetry. We will see that this has important  consequences, for example that 
   sectors with even and odd $l$ will not decouple, which will present challenges in certain cases.  
   \item
  In the second  step $\hat \gamma^{-1}$ contracts the distances in the direction $\mathbf{v}\propto e_x+e_y$ and keeps the distances perpendicular to that, so the mesh is not modified in $z$ direction. 
  \end{enumerate}
  The resulting unit cell in Fig. \ref{fig_mesh_110}c has the form of rhombic prism and 
  the mesh $P_d$ 
  is invariant only under four transformations $\hat R$ listed in Table \ref{tab_c2v_char}. There $Id$ denotes the identity, $C_n(\mathbf{V})$ denotes a rotation by $2\pi/n$ around $\mathbf{V}$, while $\sigma(\mathbf{V})$ denotes the reflection with respect to the plane perpendicular to $\mathbf{V}$. These four transformations form group 
  $C_{2v}$, which has only one-dimensional irreducible representations. For the one-dimensional irreducible representation "irrep" the transformation $\hat R$ on a vector $\mathbf{u}$ is given by the character $\chi^{irrep}(\hat R)$ 
  \begin{equation}
  \hat R ~\mathbf{u}=\chi^{irrep}(\hat R)~\mathbf{u}\ ,\quad \chi^{irrep}(\hat R)=\pm 1\ ,\qquad \mathrm {for\ 1D \ irrep}\ .
  \end{equation}
  The characters
  of the irreducible ($A_{1,2},B_{2,3}$) and reducible ($lm$) representations are given in Table \ref{tab_c2v_char} along with an example of polynomials and vectors $\mathbf{u}$   that transform according to these representations\footnote{With the change of coordinates $e_z'=\tfrac{1}{\sqrt{2}}(e_x+e_y)$, $e_y'=\tfrac{1}{\sqrt{2}}(e_x-e_y)$ and $e_x'=e_z$ our notation for $C_{2v}$ coincides with the more conventional one, but we stick to our notation as it  is more appropriate for $\mathbf{d}=e_x+e_y$. Our naming for $B_{2,3}$ agrees with \cite{feng_proc,rho_our} in $m_1=m_2$ limit, while $B_1$ does not have an analog for $m_1\not =m_2$. $B_3$ in \cite{fu} is denoted by $B_1$. }.       

\begin{table}
\begin{center}
  \begin{tabular}{c c| c c c c | c |c }
  respresent. & dim & $Id$ & $C_2(e_x+e_y)$ & $\sigma(e_x-e_y)$ & $\sigma(e_z)$ & polynom.  & vector  $\mathbf{u}$    \\                                                            
  \hline
  irred. $A_1$ & 1 & 1 & 1 & 1 & 1 & 1 , $x+y$   & $\mathbf{0}$, $e_x+e_y$ \\
  irred. $A_2$ & 1 & 1 & 1 & -1 & -1 & $(l>1)$ & $(l>1)$ \\
  irred. $B_3$ & 1 & 1 & -1 & 1 & -1 & $z$ &  $e_z$\\
  irred. $B_2$ & 1 & 1 & -1 & -1 & 1 &  $x-y$ & $e_x-e_y$ \\
\hline
  $\Gamma^{l=0}$ & 1   & 1 & 1  & 1  & 1 &  $Y_{00}$ & \\
  $\Gamma^{l=1}$ & 3   & 3 & -1 & 1  & 1 &  $Y_{10},Y_{11},Y_{1-1} $ & \\
 \hline
  \end{tabular}
  \caption{\label{tab_c2v_char} Characters $\chi(\hat R)=\sum_{i=1}^{dim} D(\hat R)_{ii}$ of representations $D$  for transformations $\hat R\in C_{2v}$ (with principal axis $e_x+e_y$), that leave the mesh $P_d$  in Fig. \ref{fig_mesh_110} for $\mathbf{d}=e_x+e_y$ invariant. 
Representations $A_{1,2}$ and $B_{2,3}$ are irreducible while the representation  $\Gamma^{l=1}$ is  reducible. 
Example of polynomials and vectors $\mathbf{u}$  that transform according to these representations are given on the right. }
\end{center}
  \end{table}

The functions $Y_{lm}$ (\ref{Y_tr}) and also the solutions $G_{lm}^{\mathbf{d}}$ (\ref{G_tr}) form a representation $\Gamma^{(l)}$ for transformations $\hat R\in C_{2v}$, but the $2 l+1$ dimensional representation $\Gamma^{(l)}$ with $m=-l,..,l$ is in general reducible. We will need the number of times  ($N^{irrep}$) that irreducible representation ``irrep'' enters in $\Gamma^{(l)}$  \cite{dawber}
\begin{equation}
\label{decompose}
N^{irrep}=\frac{1}{g}\sum_{\hat R} \chi^{irrep}(\hat R)^*~\chi^{(l)}(\hat R)\ ,
\end{equation}
where $g$ is the number of elements $\hat R$ of the group $G$, while the characters of irreducible representations  $\chi^{irrep}(\hat R)$ and reducible representations $\chi^{(l)}(\hat R)$ are given in Table \ref{tab_c2v_char}. The resulting decomposition is\footnote{The decomposition agrees with  \cite{fu}, but there $B_3$ is denoted by $B_1$.} 
\begin{align}
\label{decompose_110}
\Gamma^{(0)}&=A_1\\
\Gamma^{(1)}&=A_1\oplus B_3 \oplus B_2\nonumber\\
\Gamma^{(2)}&=2A_1\oplus A_2 \oplus B_3 \oplus B_2\nonumber
\end{align}
This indicates that the solutions (and also interpolators) that transform according to the listed irreps will contain the following partial waves
\begin{equation}
 B_3:\ l=1,~2,..\qquad B_2:\ l=1,~2,..\qquad A_1:\ l=0,~1,~2,.. \qquad A_2:\ l=2,..
\end{equation}
 so $B_3$ or $B_2$ in CMF will couple to $l=1$, but not to $l=0$. These two representations therefore provide a rather clean possibility to extract $l=1$ phase shift if partial waves with $l>1$ can be neglected. The interpolators that transform according to irrep $A_1$ in CMF will couple to both $l=0$ and $l=1$, and there is unfortunately no irreducible representation which would couple only to $l=0$. This will present a serious challenge for a reliable extraction of $l=0$ phase shift in simulations with non-zero total momentum, as will be discussed in more detail later on. 

The mixing between the even and odd $l$ occurs since inversion is not an element of the group $C_{2v}$ (see Table \ref{tab_c2v_char}).  
We emphasize  that this mixing is not present 
for the scattering of  particles with $m_1=m_2$ and total momentum $\mathbf{d}=e_x+e_y$ since inversion is element of $D_{2h}$:  
 $A^+$ contains only the waves $l=0,2$ and $l\geq 4$, while $B_{1,2,3}^-$ contains only $l=1$ and $l\geq 3$ \cite{feng_proc}. This mixing is also not present for the scattering of particles with $m_1\not =m_2$ and total momentum $\mathbf{P}=0$, where $A_1^+$ contains only $l=0,4,..$ and $T_1^-$ contains only $l=1,3,..$ \cite{luscher_npb}. 
  
  \subsubsection{Values of $Z_{lm}^{\mathbf{d}}$ as consequences of the symmetries for $\mathbf{d}=e_x+e_y$}\label{symmetries_c2v}

  The transformations $\hat R\in C_{2v}$ leave the mesh $P_d$ invariant, so the sum over $\mathbf{r}\in P_d$ in $Z_{lm}$ (\ref{Z_def}) can be replaced by the sum over $\mathbf{r}'=\hat R \mathbf{r}$  \begin{align}
\label{Z_sym}
Z_{lm}^{\mathbf{d}}(q^2)&=\sum_{\mathbf{r}}\frac{{\cal Y}_{lm}(\mathbf{r})}{(\mathbf{r}^2-q^2)^s}= \sum_{\mathbf{r}'=\hat R\mathbf{r}}\frac{|\mathbf{r}'|^l~{Y}_{lm}(\mathbf{r}')}{(\mathbf{r}'^2-q^2)^s}=\sum_{\mathbf{r}}\frac{\mathbf{r}^l~Y_{lm}(R\mathbf{r})}{(\mathbf{r}^2-q^2)^s}\qquad \mathrm{since}\ |\mathbf{r}'|=|\hat R\mathbf{r}|=|\mathbf{r}|
%=\sum_{\vec r}\frac{1}{(\vec r^2-q^2)^s}\sum_{m'} D_{mm'}(R)Y_{lm'}(\vec r)=\sum_{\vec m'}~D_{mm'}(R)~Z_{lm'}(q^2),\nonumber
\end{align}
which will have important consequences for some $\hat R$. Now we can list the properties of   $Z^{\mathbf{d}}_{lm}$ for $\mathbf{d}=e_x+e_y$:  

\begin{itemize}
\item $Z^{\mathbf{d}}_{l~-m}=(-1)^m(Z^{\mathbf{d}}_{lm})^*$ which is the consequence of the analogous relation for $Y_{lm}$.
\item $Z^{\mathbf{d}}_{lm}=i^m (Z^{\mathbf{d}}_{lm})^*$  since $\hat R=\sigma(e_x-e_y)\in C_{2v}$:

The reflection $\hat R=\sigma(e_x-e_y)$ with respect to the plane  perpendicular to $e_x-e_y$ transforms 
$x\to y,~y\to x,~ z\to z$ or equivalently $\theta\to\theta,\ e^{i\varphi}\propto x+iy\to y+ix=i(x-i y)\propto ie^{-i\varphi}$. This transforms $Y_{lm}\equiv f(\theta)~(e^{i\varphi})^m~\to~ f(\theta)( ie^{-i\varphi})^m=f(\theta)i^m e^{-im\varphi}=i^m(Y_{lm})^*$ and  then (\ref{Z_sym}) leads to $Z_{lm}^{\mathbf{d}}= i^m  (Z^{\mathbf{d}}_{lm})^*$. 

Rewriting $Z^{\mathbf{d}}_{lm}=Re(Z^{\mathbf{d}}_{lm})+i~Im(Z^{\mathbf{d}}_{lm})$ this leads to $Im(Z^{\mathbf{d}}_{lm})=0$ for $m=0,4,..$,  $Re(Z^{\mathbf{d}}_{lm})=Im(Z^{\mathbf{d}}_{lm})$ for $m=1,5,..$, $Re(Z^{\mathbf{d}}_{lm})=0$ for $m=2,6,..$ and $Re(Z^{\mathbf{d}}_{lm})=-Im(Z^{\mathbf{d}}_{lm})$ for $l=3,7,..$, independent of the value of $l$. 

This property of $Z^{\mathbf{d}}_{lm}$ holds for any case where the mesh $P_d$ is symmetric under $\hat R=\sigma(e_x-e_y)$, which is true for all $\mathbf{d}=0,~e_x+e_y,~e_z$ in case of degenerate or non-degenerate masses.
\item $Z_{lm}^{\mathbf{d}}=0$ for $l-m=odd$ since $\sigma(e_z)\in C_{2v}$:

The reflection $\hat R=\sigma(e_z)$ with respect $xy$ plane transforms 
$\theta\to \pi-\theta$ and $\varphi\to\varphi$, while $Y_{lm}(\pi-\theta,\varphi)=(-1)^{l-m}Y_{lm}(\theta,\varphi)$ so (\ref{Z_sym}) leads to  $Z_{lm}^{\mathbf{d}}=(-1)^{l-m}Z_{lm}^{\mathbf{d}} $. 
\item Note that if inversion $I$ would be in $G$ then $Z^{\mathbf{d}}_{lm}=(-1)^l Z^{\mathbf{d}}_{lm}$ or $Z^{\mathbf{d}}_{lm}=0$ for odd $l$. This would decouple parts of $M$ for even and odd $l$. In the present case $m_1\not = m_2$ and $\mathbf {d}\not =0$,  so $I$ is not element of $G$ and $Z_{lm}^{\mathbf{d}}$ is not zero in general for odd $l$. 
\end{itemize}
We verified all the above relations to be true also with the explicit numerical evaluation of $Z^{\mathbf{d}}_{lm}$ using the expression in Appendix A. For example $Z^{\mathbf{d}}_{11}(m_1\not =m_2)\not =0 $ and has equal real and imaginary part, while $Z^{\mathbf{d}}_{11}(m_1 =m_2)=0 $ as required by the symmetries of $D_{2h}$ for the mass-degenerate case \cite{feng_proc}. 

\subsubsection{The matrix ${\cal M}_{lm,l'm'}^{\mathbf{d}}$ for $\mathbf{d}=e_x+e_y$}

The above relations for $Z_{lm}^{\mathbf{d}}$ or equivalently $w_{lm}$ (\ref{w}) simplify the general matrix ${\cal M}$ (\ref{M_gen}) to 
\begin{equation}
\label{M_110}
M={\cal M}_{lm,l'm'}^{\mathbf{d}} = \bordermatrix{~ & 00        & 10                    & 11                       & 1-1\cr
  00 & w_{00} & 0 & i \sqrt{3} w_{11} & -\sqrt{3} w_{11} \cr
  10 & 0      & w_{00}+2w_{20} & 0 & 0 \cr
  11 & - \sqrt{3}w_{11} & 0 & w_{00}-w_{20} & \sqrt{6} w_{22} \cr
 1-1 & i \sqrt{3} w_{11} & 0 & -\sqrt{6}w_{22} & w_{00}-w_{20} \cr}
\end{equation}
since $iw_{1-1}=-iw_{11}^*=-w_{11}$ due to $Re(w_{11})=Im(w_{11})$, and $w_{2-2}=w_{22}^*=-w_{22}$ due to $Re(w_{22})=0$, while $w_{00}$ and $w_{20}$ are real.  

\subsubsection{Phase shift relations for $\mathbf{d}=e_x+e_y$} 

 The phase shift relations are obtained from the determinant condition (\ref{det}), which gets simplified when $M$ (\ref{M_110}) is written in the basis that renders it block-diagonal. \\

{\bf Extracting $P$-wave phase shift from irreps $B_{3}$ or $B_2$ }

\vspace{0.2cm}

The part $Y_{10}\propto z $ already presents a separate block, which transforms according to $B_3$: $z$ gets multiplied by $-1$ for reflection with respect to $xy$ plane and rotation around $e_x+e_y$ and stays invariant for other two transformations (see Table \ref{tab_c2v_char}). The determinant condition for this $1\times 1$ block requires   $det[e^{2i\delta_1}(M^{B_3}\!-\!i)-(M^{B_3}\!+\!i)]=0$ or equivalently $\tan(\delta_1)=1/M^{B_3}$ with $M^{B_3}=w_{00}+2w_{20}$, so
\begin{equation}
\label{ph_110_B3}
\mathbf{d}=e_x+e_y~, \ B_3\ \mathrm{of}\ C_{2v}  \ :
\quad \tan\delta_1(p^*)=\frac{\pi^{3/2}~\gamma ~q}{Z_{00}^{\mathbf{d}}(1;q^2)+\tfrac{2}{\sqrt{5}}~q^{-2}~Z_{20}^{\mathbf{d}}(1;q^2)} \ .  
 \end{equation}
This is the final relation that allows the determination of $P$-wave phase shift $\delta_1(p^*)$ from the energy $E$ of two particles with total momentum $\mathbf{P}=\tfrac{2\pi}{L}(e_x+e_y)$ when one uses the interpolators that transform according to $B_3$. 

The eigenvectors of the remaining $3\times 3$ matrix $M$ (\ref{M_110}) reveal that one simple eigenvector is $\mathbf{u}^{B_2}=\tfrac{1}{\sqrt{2}} (-iY_{11}+Y_{1-1})\propto x-y$ which transforms according to $B_2$ (see Table \ref{tab_c2v_char}). The corresponding eigenvalue is $M^{B_2}=w_{00}-w_{20}-\sqrt{6} Im(w_{22})$ and the determinant condition (\ref{det}) for this $1\times 1$ block gives $\tan(\delta_1)=1/M^{B_3}$ , or
\begin{equation}
\label{ph_110_B2}
 \mathbf{d}=e_x+e_y~, \ B_2\ \mathrm{of}\ C_{2v} ~ :
 \quad \tan\delta_1(p^*)=\frac{\pi^{3/2}~\gamma ~q}{Z_{00}^{\mathbf{d}}(1;q^2)-\tfrac{1}{\sqrt{5}}~q^{-2}~Z_{20}^{\mathbf{d}}(1;q^2)-\tfrac{\sqrt{6}}{\sqrt{5}}~q^{-2}~Im[Z_{22}^{\mathbf{d}}(1;q^2)]}.  
 \end{equation}
This is another relation that allows determination of $\delta_1(p^*)$ when using interpolators in irrep $B_2$. 

The phase shift relations for irreps $B_3$ (\ref{ph_110_B3}) and $B_2$ (\ref{ph_110_B2}) agree\footnote{Note that $Z_{lm}^{\mathbf{d}}$ in \cite{feng_proc,rho_our} is complex conjugate of our $Z_{lm}^{\mathbf{d}}$ (\ref{Z_def}).}  with the expressions in \cite{feng_proc,rho_our} for the case $m_1=m_2$. The remaining representations, discussed bellow, have different role in the $m_1=m_2$ and $m_1\not =m_2$ cases. \\

{\bf Problems and  strategies for extracting $S$-wave phase shift from irrep $A_1$}

\vspace{0.2cm}

The remaining $2\times 2$ matrix ${M}$ can not be reduced further and spans the space in the basis  of vectors $\mathbf{u}^{A_1}_1=\tfrac{1}{\sqrt{2}}(Y_{11}-iY_{1-1})\propto x+y$ and $\mathbf{u}^{A_1}_2=Y_{00}\propto 1$, which are perpendicular to the vectors $\mathbf{u}^{B_3}=Y_{10}$ and $\mathbf{v}^{B_2}$  above. The vectors $x+y$ and $1$ both remain invariant under all four $\hat R\in C_{2v}$ and belong to $A_1$ irreducible representation (see Table \ref{tab_c2v_char}). The remaining $2\times 2$ block in the   basis  $\mathbf{u}^{A_1}_{1,2}$ is contained in 
\begin{equation}
\label{M_110_block}
 M_{ab}^{B} = \bordermatrix{~ & Y_{00}        & \tfrac{Y_{11}-iY_{1-1}}{\sqrt{2}}     & Y_{10}        & \tfrac{-iY_{11}+Y_{1-1}}{\sqrt{2}}\cr
  Y_{00}                               & w_{00} & i\sqrt{6}w_{11} & 0 & 0 \cr
  \tfrac{Y_{11}-iY_{1-1}}{\sqrt{2}} & -i \sqrt{6}w_{11}^*   & w_{00}-w_{20}+\sqrt{6} Im(w_{22}) & 0 & 0 \cr
  Y_{10}                               & 0 & 0 & w_{00}+2w_{20} & 0 \cr
  \tfrac{-iY_{11}+Y_{1-1}}{\sqrt{2}}& 0 & 0 & 0 & w_{00}-w_{20}-\sqrt{6} Im(w_{22})\cr}
\end{equation}
with $a,b=0,..,3$. We kept the other two $1\times 1$ blocks for completeness, so  $M^{B}$ represents the desired block-diagonal form of $M$ in the basis $\mathbf{u}_{1,2}^{A_1},\mathbf{u}^{B_3},\mathbf{u}^{B_2}$  and the superscript ``B'' refers to block-diagonal. The determinant condition (\ref{det}) is equivalent to determinant conditions for three separate blocks and two of those  were  already written in (\ref{ph_110_B3},\ref{ph_110_B2}).
The determinant condition for $2\times 2$ block leads to the relation
\begin{align}
\label{ph_110_A1}
&\mathbf{d}=e_x+e_y~,  \  A_1\ \mathrm{of}\ C_{2v}\ :\\
&[e^{2i\delta_0(p^*)}(M^{B}_{00}-i)-(M^{B}_{00}+i)][e^{2i\delta_1(p^*)}(M^{B}_{11}-i)-(M^{B}_{11}+i)]=|M_{10}|^2(e^{2i\delta_0(p^*)}-1)(e^{2i\delta_1(p^*)}-1).\nonumber 
\end{align}

\vspace{0.2cm}

Among  three phase shift relations (\ref{ph_110_B3},\ref{ph_110_B2},\ref{ph_110_A1}), the $S$-wave phase shift $\delta_0$ enters only in (\ref{ph_110_A1}), so let us discuss the problems and possible strategy for extracting $\delta_0$ from (\ref{ph_110_A1}). 
Suppose we determine the energy level $E$ using interpolators that transform according to representation $A_1$. The values of $E,~A,~\mathbf{d}$ provide the values of $M^{B}_{00,11,10}(q^2)$ in (\ref{ph_110_A1}) at corresponding $q=p^*L/2\pi$ (\ref{helmholtz2}), so the relation (\ref{ph_110_A1})  presents one equation with two unknowns: $\delta_0(p^*)$ and $\delta_1(p^*)$. In order to determine $\delta_0(p^*)$ from (\ref{ph_110_A1})
\begin{align}
\label{delta_0}
&\mathbf{d}=e_x+e_y~,  \  A_1\ \mathrm{of}\ C_{2v}\ :\\
&\tan\delta_0(p^*)=\frac{1-\tan\delta_1(p^*)~[w_{00}-w_{20}+\sqrt{6}~Im(w_{22})]}{w_{00}-\tan\delta_1(p^*)~[w_{00}^2-w_{00}w_{20}+\sqrt{6}w_{00}~Im(w_{22})-12 (Re~w_{11})^2]}\nonumber
\end{align}
one would need to know the value of $\delta_1(p^*)$ at given $p^*$ (\ref{helmholtz2}). The representations $B_3$ or $B_2$ allow determination of 
$\delta_1(\tilde p^*)$, but the problem is that they will in general fix $\delta_1$ at some other value of $\tilde p^*$, which is related to the energy $\tilde E$ measured for the case of representations $B_3$ or $B_2$. Since  $\delta_1(p^*)$ in (\ref{ph_110_A1}) is needed at $p^*$, $\delta_1(\tilde p^*)$ can not be simply used if $\tilde p^*\not = p^*$. 

This indicates that the reliable extraction of $S$-wave phase shift will be challenging, since there is no irreducible representation that would contain only $\delta_0$ but not $\delta_1$.    We envisage several possible strategies to estimate the value of $\delta_0$, which might be used in the pioneering simulations along these directions, but it is clear that some of these strategies  are not rigorous:
\begin{enumerate}
\item If there exists a region of $p^*$ where $\delta_1(p^*)$ is known to be negligible, 
the condition (\ref{ph_110_A1},\ref{delta_0}) recovers the standard form \cite{savage,fu} $\tan(\delta_0)=1/w_{00}$ or
\begin{equation}
\label{ph_110_A1_neglect}
 \mathbf{d}=e_x+e_y~,  \  A_1\ \mathrm{of}\ C_{2v}\ :\qquad \tan\delta_0(p^*)=\frac{\pi^{3/2}~\gamma ~q}{Z_{00}^{\mathbf{d}}(1;q^2)}  \qquad \mathrm{if}\  \delta_1(p^*)\ll \delta_0(p^*)~
 \end{equation}
 and allows the determination of $\delta_0$ for that $p^*$ \cite{fu}. That may be for example possible at small $p^*$, where higher partial waves are generally suppressed, but $p^*$ has to be away from  any nearby $P$-wave resonance 
where $\delta_1(p^*)\simeq \pi/2$ is not small.
 \item 
If $\delta_1(p^*)$ is not negligible, and its $p^*$ dependence is expected to be mild, then one can estimate  $\delta_1(p^*)$ needed in (\ref{ph_110_A1}) from 
$\delta_1(\tilde p^*)$ using the interpolation between $\tilde p^*$ and $p^*$. 
In this case $\delta_1(\tilde p^*)$ has to be determined for several $\tilde p^*$ using several  representations and several  total momenta $\mathbf{P}$ (for example $B_{3,2}$ at $\mathbf{d}=e_x+e_y$). 
\item 
In the continuum limit, one expects that the energy levels $E$ determined using different irreducible representations will agree for a physical state with a given total momentum $\mathbf{P}$ and given $l$. In the past this (near) degeneracy across different representations  served to determine $l$ (or $J$) of the resulting states. In our case of interest, we expect $E_{l=1}^{A_1}=E_{l=1}^{B_3}=E_{l=1}^{B_2}$ for $\mathbf{d}=e_x+e_y$ in the continuum limit, so the resulting $p^*$ for $l=1$ state will be the same for all three irreps. This allows the determination of $\delta_1(p^*)$ at the desired $p^*$ from $B_{2,3}$;  inserting that to the phase shift relation for $A_1$ (\ref{ph_110_A1}) will finally allow the extraction of $\delta_0(p^*)$. In practice, the equality between $E$ or $p^*$ from different representations will be slightly spoiled by the discretization errors. This procedure would still give a relatively reliable  estimate of $\delta_0(p^*)$ if $\delta_1$ modestly depends on $p^*$, i.e. if there is no close by narrow $P$-wave resonance.
\end{enumerate} 

We expect that more a reliable extraction for $S$-wave phase shift of two particles with $m_1\not =m_2$ needs to be performed using a simulation with $\mathbf{P}=0$. There representation $A_1$ mixes $\delta_0$ only with $\delta_4$ and even higher partial waves, which can be safely neglected. The drawback of sticking to a single total momentum $\mathbf{P}=0$ is that one needs to perform simulations at several lattice sizes $L$ in order to determine $\delta_0(p^*)$ as several values of $p^*$.

\subsubsection{Quark-antiquark and meson-meson interpolators for $\mathbf{d}=e_x+e_y$}

In order to extract the phase shifts, one needs to simulate the two-particle system on a finite lattice and determine its energy $E$ in the lattice frame. To create the two-particle state, one may use the corresponding two-particle interpolator or the  quark-antiquark interpolator, that couples well to the two-particle state or the resonance that appears in this channel. The interpolators are written down in the lattice frame, but they have to transform according to the desired irreducible representation after the transformation to  the CMF. In this section we write down some simple examples of such interpolators with this property, that may be used in the lattice simulations.

For concreteness, our two-particle interpolators refer to two pseudoscalar mesons $P_1P_2$ with masses $m_1$ and $m_2$, since pseudoscalar mesons are often stable against the strong decay and therefore their scattering is most interesting phenomenologically. In the continuum, the scattering state  therefore carries $J^P=0^+$ for $l=0$ and $J^P=1^-$ for $l=1$ in our examples. We write down examples of interpolators for this case, but this can be generalized to the scattering of other types of particles. 

All interpolators will be expressed in terms of the currents ($i=x,y,z$) 
\begin{equation}
\label{currents}
 V_{i}(\mathbf{p})\equiv \sum_{\mathbf{x}} e^{i \mathbf{p x}}\bar q(\mathbf{x})\gamma_i q'(\mathbf{x})   \quad P(\mathbf{p})\equiv \sum_{\mathbf{x}} e^{i \mathbf{p x}}\bar q(\mathbf{x})\gamma_5 q'(\mathbf{x})\quad S(\mathbf{p})\equiv \sum_{\mathbf{x}} e^{i \mathbf{p x}}\bar q(\mathbf{x}) q'(\mathbf{x})~,
 \end{equation}
 where one can replace the choices of the $\gamma$ matrix in the current with any combination of $\gamma$ matrices and covariant derivatives  that gives the same transformation property of the current. In this subsection we write all the momenta in units of $2\pi/L$.

Examples of  the quark-antiquark interpolators in lattice frame that transform according to $B_2,~ B_3$ and $A_1$ in CMF  are
\begin{align}
{\cal O}^{\bar qq}_{B_2}&=V_x(e_x+e_y)-V_y(e_x+e_y)\nonumber\\
{\cal O}^{\bar qq}_{B_3}&=V_z(e_x+e_y)\nonumber\\
({\cal O}^{\bar qq}_{A_1})^I&=V_x(e_x+e_y)+V_y(e_x+e_y)\nonumber\\
({\cal O}^{\bar qq}_{A_1})^{II}&=S(e_x+e_y)~.\nonumber
\end{align}
More general quark-antiquark interpolators with non-zero momentum are constructed  in \cite{fleming_qq,dudek_nonzero}.

Let us consider ${\cal O}^{\bar qq}$  transformations on the example of ${\cal O}^{\bar qq}_{B_3}$, which becomes $({\cal O}^{\bar qq}_{B_3})_{CMF}=V_z(0)$ after the boost to CMF. The boost   along $\mathbf{d}=e_x+e_y$  does not modify its polarisation  (\ref{hat_gam}).  Interpolator $V_z(0)$ in CMF transforms like $e_z$, so according to $B_3$ representation in Table \ref{tab_c2v_char}.

\vspace{0.2cm}

The two-particle interpolators are linear combinations  with  momentum choices   $\mathbf{p}_{1j}$ and $\mathbf{p}_{2j}$
\begin{align}
{\cal O}^{P_1P_2}_{irrep} &=\sum_{j=a,b,..} P_1(\mathbf{p}_{1j})~P_2(\mathbf{p}_{2j})\\
({\cal O}^{P_1P_2}_{irrep})_{CMF}& =\sum_j P_1(\mathbf{p}_{j}^*)~P_2(-\mathbf{p}_{j}^*) \ ,\qquad \mathbf{p}_{j}^*=\mathbf{p}_{1j}^*=-\mathbf{p}_{2j}^*=\hat \gamma^{-1}[\mathbf{p}_1-\tfrac{1}{2}A\mathbf{P}]\nonumber\\
\hat R ({\cal O}^{P_1P_2}_{irrep})_{CMF} \hat R^{-1} &= \sum_j P_1(\hat R \mathbf{p}_{j}^*)~P_2(-\hat R \mathbf{p}_{j}^*) \stackrel {\mathrm{1D\ irrep}}{ =} \chi^{irrep}(\hat R)~ ({\cal O}^{P_1P_2}_{irrep})_{CMF}\nonumber
\end{align}
 such that they transform according to given irrep in CMF, where  momenta $\mathbf{p}_{1j}^*=-\mathbf{p}_{2j}^*=\mathbf{p}_j^*$ are given by (\ref{pstar}). 
  Examples of interpolators  with this property are 
\begin{align}
\label{interpolators_c2v}
({\cal O}^{P_1P_2}_{B_{2}})^I &=P_1(e_x)P_2(e_y)-P_1(e_y)P_2(e_x)\\
({\cal O}^{P_1P_2}_{B_{2}})^{II} &=P_1(e_x+e_z)P_2(e_y-e_z)-P_1(e_y+e_z)P_2(e_x-e_z)+  \{ e_z\leftrightarrow -e_z\}\nonumber\\
&~\nonumber\\
({\cal O}^{P_1P_2}_{B_{3}})^I &=P_1(e_x+e_y+e_z)P_2(-e_z)-P_1(e_x+e_y-e_z)P_2(e_z)\nonumber\\
({\cal O}^{P_1P_2}_{B_{3}})^{II} &=P_1(e_x+e_z)P_2(e_y-e_z)+P_1(e_y+e_z)P_2(e_x-e_z)-  \{ e_z\leftrightarrow -e_z\}\nonumber\\
&~\nonumber\\
({\cal O}^{P_1P_2}_{A_{1}})^I &=P_1(e_x+e_y)P_2(0)\nonumber\\
({\cal O}^{P_1P_2}_{A_{1}})^{II} &=P_1(e_x)P_2(e_y)+P_1(e_y)P_2(e_x)\nonumber\\
({\cal O}^{P_1P_2}_{A_{1}})^{III} &=P_1(e_x+e_z)P_2(e_y-e_z)+P_1(e_y+e_z)P_2(e_x-e_z)+  \{ e_z\leftrightarrow -e_z\}\nonumber\\
({\cal O}^{P_1P_2}_{A_{1}})^{IV} &=P_1(e_x+e_y+e_z)P_2(-e_z)+P_1(e_x+e_y-e_z)P_2(e_z)\nonumber
\end{align}
and the analogous interpolators where flavors of $P_1$ and $P_2$ are interchanged. The  interpolators  were obtained using the projection operator $\tfrac{dim^{irrep}}{g} \sum_{\hat R} \chi^{irrep} (\hat R) ~T(\hat R)$ and  we list all $P_1P_2$  interpolators that have $p_{1}\leq \sqrt{3} ~\tfrac{2\pi}{L}$ and $p_{2}\leq \sqrt{3} ~ \tfrac{2\pi}{L}$.  
More general hadron-hadron interpolators  are considered in \cite{fleming_MM}.

\vspace{0.3cm}

The  correct transformation properties of the interpolators  (\ref{interpolators_c2v}) can be easily demonstrated if the momenta 
$\mathbf{p}_{1j}$ and $\mathbf{p}_{2j}$ are written as a sum of  a vector parallel to $\mathbf{d}$ and a  vector $\mathbf{u}\perp \mathbf{d}$.  
Let us demonstrate that $ ({\cal O}^{P_1P_2}_{B_{3}})^{I} =P_1(\mathbf{d}+\mathbf{u})P_2(-\mathbf{u})-P_1(\mathbf{d}-\mathbf{u})P_2(\mathbf{u})$ with $\mathbf{u}= e_z \perp \mathbf{d}$ transforms according to $B_3$. The momenta of $P_1$ in CMF are (\ref{pstar})
\begin{align}
\mathbf{p}_{1a} &=\mathbf{d}+\mathbf{u} \ :\quad \mathbf{p}_a^*=\hat \gamma^{-1}(\mathbf{p}_{1a}-\tfrac{1}{2}A\mathbf{P})=\hat \gamma^{-1}(\mathbf{d}+\mathbf{u}-\tfrac{1}{2}A\mathbf{d})=\gamma^{-1}(1-\tfrac{1}{2}A)\mathbf{d}+\mathbf{u}=c\mathbf{d}+\mathbf{u}\nonumber\\
\mathbf{p}_{1b} &=\mathbf{d}-\mathbf{u} \ :\quad \mathbf{p}_b^*=c\mathbf{d}-\mathbf{u}\qquad c\equiv \gamma^{-1}(1-\tfrac{1}{2}A)~,
 \end{align}
 while the interpolator in CMF is    
  \begin{equation}
 \label{O_PP_CMF}
 ({\cal O}^{P_1P_2}_{B_{3}})^{I}_{CMF} =P_1(c\mathbf{d}+\mathbf{u})P_2(-c\mathbf{d}-\mathbf{u})-P_1(c\mathbf{d}-\mathbf{u})P_2(-c\mathbf{d}+\mathbf{u}) \equiv {\cal O}(\mathbf{d},\mathbf{u})~.
 \end{equation}
 Due to the following properties it transforms according to 1D irrep $B_{3}$
 \begin{align}
  {\cal O}(\mathbf{d},-\mathbf{u})&=-  {\cal O}(\mathbf{d},\mathbf{u})\quad \mathrm{or}\quad   {\cal O}(\mathbf{d},\hat R\mathbf{u})=\chi^{irrep.}(\hat R) ~ {\cal O}(\mathbf{d},\mathbf{u})\\
  \hat R {\cal O}(\mathbf{d},\mathbf{u})\hat R^{-1}&= {\cal O}(\hat R \mathbf{d}, \hat R \mathbf{u})= {\cal O}(\mathbf{d}, \chi^{irrep}(\hat R) \mathbf{u})=\chi^{irrep}(\hat R)~ {\cal O}(\mathbf{d},  \mathbf{u})\nonumber
 \end{align}
 since $\hat R\in C_{2v}$ leave $\mathbf{d}=e_x+e_y$ unaffected, while $\hat R\mathbf{u}=\chi^{irrep}(\hat R)\mathbf{u}=\pm \mathbf{u}$ for $\mathbf{u}=e_z$, as listed in Table \ref{tab_c2v_char}. The procedure for treating transformations of other interpolators in (\ref{interpolators_c2v}) is analogous.

\subsection{$\mathbf{P}= (2\pi/L) e_z$ and consequences of $C_{4v}$ symmetries}
 
Equipped with the knowledge how to handle the momentum $\mathbf{d}=e_x+e_y$, which was more general, one can easily consider $\mathbf{d}=e_z$, which has more symmetry transformations. 

\begin{figure}[bt]
\begin{center}
\includegraphics*[width=0.2\textwidth,clip]{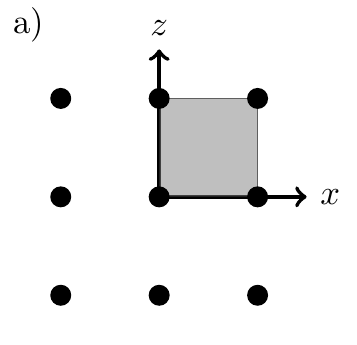} $\quad$
\includegraphics*[width=0.2\textwidth,clip]{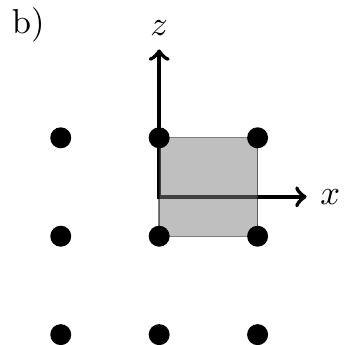} $\quad$
\includegraphics*[width=0.2\textwidth,clip]{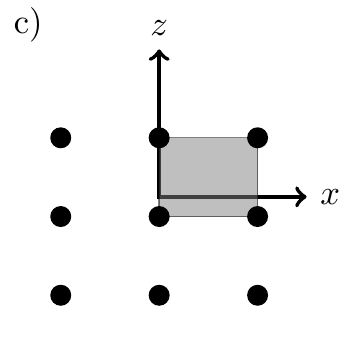}
\end{center}
\caption{The mesh $P_d$ (\ref{P_d}) for $\mathbf{d}=e_z$ and $m_1\not =m_2$ ($A\not =1$) is plotted in (c), while (a) and (b) show the steps how to get it.  }\label{fig_mesh_001}
\end{figure}

The mesh $P_d$ (\ref{P_d}) in Fig. \ref{fig_mesh_001}c is obtained from $\mathbf{n}\in Z^3$ by the shift $-\tfrac{1}{2}Ae_z$ and the inversion is lost at this stage. Then the lengths in $e_z$ direction are contracted due to  $\hat \gamma^{-1}$. 

The transformations $\hat R$ that leave this mesh invariant are given in  Table \ref{tab_c4v_char} and they form the group $C_{4v}$. It has four 1D-irreps $A_{1,2},~B_{1,2}$ and one 2D-irrep $E$, where only 
$A_1$ and $E$ appear for   $l=0,1$ of our interest according to (\ref{decompose}) \cite{fu}
\begin{align}
\label{decompose_001}
\Gamma^{(0)}&=A_1\\
\Gamma^{(1)}&=A_1\oplus E\nonumber\\
\Gamma^{(2)}&=A_1\oplus B_1 \oplus B_2 \oplus E\ .\nonumber
\end{align}
So the interpolators that transform according to the listed irreps contain the following partial waves:
\begin{equation}
E:\ l=1,~2,.. \qquad  A_1:\ l=0,~1,~2,.. \qquad B_1:\ l=2,..\qquad B_2:\ l=2,.. 
\end{equation}

We note that the mixing between even and odd $l$ is not present 
for the scattering of particles with $m_1=m_2$ and total momentum $\mathbf{d}=e_z$ since inversion is an element of $D_{4h}$:  
$A_1^+$  contains only the waves $l=0,2$ and $l\geq 4$, while $E^-$ and $A_2^-$ contain only $l=1$ and $l\geq 3$ \cite{RG}. This mixing is also not present for the scattering of particles with $m_1\not =m_2$ and total momentum $\mathbf{P}=0$.

\begin{table}
\begin{center}
  \begin{tabular}{c c| c c c c  c | c |c }
  respresent. & dim & Id & $ C_4(e_z)$  & $C_2(e_z)$ & $\sigma(e_x)$ & $\sigma(e_x+e_y)$ & polynom.  & vector  $\mathbf{u}$    \\   
 &  &  & $ C_4^{-1}(e_z)$  & & $\sigma(e_y)$ & $\sigma(e_x-e_y)$ &   &     \\   
    \hline
  irred. $A_1$ & 1 & 1 & 1 & 1 & 1& 1 & 1 , $z$   & $\mathbf{0}$, $e_z$ \\
  irred. $E$ &    2 & 2 & 0 & -2 & 0 & 0        &  $x,y$ or $Y_{11},Y_{1-1}$& $e_x$, $e_y$ \\
\hline
  $\Gamma^{l=0}$ & 1   & 1 & 1  & 1  & 1 &    1 & $Y_{00}$ & \\
  $\Gamma^{l=1}$ & 3   & 3 & 1 & -1  & 1 &   1 & $Y_{10},Y_{11},Y_{1-1} $ & \\
 \hline
  \end{tabular}
  \caption{\label{tab_c4v_char} Characters   for transformations $R\in C_{2v}$ (with principal axis $e_z$), that leave the mesh $P_d$ for $d=e_z$ in Fig. \ref{fig_mesh_001} invariant. In addition to irreps $A_1$ and $E$, $C_{4v}$ has also $A_2$ and $B_{1,2}$ but they do not appear for $l=0,1$ so we omit them.  
Example of simple objects that transform according to these representations are given on the right. }
\end{center}
  \end{table}
  
  \subsubsection{Values of $Z_{lm}^{\mathbf{d}}$ as consequences of the symmetries for $\mathbf{d}=e_z$}
  
  By comparing to the case of $\mathbf{d}=e_x+e_y$, we find that some relations are still valid, 
  some are not valid  and there are some additional ones due to the additional elements $\hat R\in C_{4v}$:
  \begin{itemize}
  \item $Z^{\mathbf{d}}_{l~-m}=(-1)^m(Z^{\mathbf{d}}_{lm})^*$ since $Y_{l-m}=(-1)^m Y_{lm}^*$.
\item $Z^{\mathbf{d}}_{lm}=i^m (Z^{\mathbf{d}}_{lm})^*$  since $\sigma(e_x-e_y)\in C_{4v}$. 
The consequences regarding specific values of $m$ are listed in the corresponding section for $\mathbf{d}=e_x+e_y$. 
%\item $Z_{lm}^{\mathbf{d}}=0$ for $m=odd$ since $C_2(e_z)\in C_{4v}$:

 %$C_2(e_z)$ transforms $\theta\to\theta,\ \varphi \to \varphi+\pi$, so $Y^{\mathbf{d}}_{lm}\equiv f(\theta)~e^{im\varphi}~\to~f(\theta)~e^{im(\varphi+\pi)}=f(\theta) (e^{i\pi})^m e^{im\varphi}$, so $Z^{\mathbf{d}}_{lm}=(-1)^m Z^{\mathbf{d}}_{lm}$.
\item $Z^{\mathbf{d}}_{lm}=0$ if $m\not=0,4,8,...$    since $C_4(e_z)\in C_{4v}$:

$C_4(e_z)$ transforms $\theta\to\theta,\ \varphi \to \varphi+\tfrac{1}{2}\pi$, so $Y^{\mathbf{d}}_{lm}\equiv f(\theta)~e^{im\varphi}~\to~f(\theta)~e^{im(\varphi+\tfrac{1}{2}\pi)}=f(\theta) (e^{i\pi/2})^m e^{im\varphi}$, so $Z^{\mathbf{d}}_{lm}=(i)^m Z^{\mathbf{d}}_{lm}$.
\item $Z_{lm}^{\mathbf{d}}$ is not zero  for $l-m=odd$ in general since $\sigma(e_z)\not \in C_{4v}$ (see Fig. \ref{fig_mesh_001}). 
\item  $Z^{\mathbf{d}}_{lm}$ is not zero for $l=odd$ in general, since $I\not \in C_{4v}$:

Therefore the matrix $M$ will not be decomposed into sectors with even and odd $l$,  so $\delta_0$ and $\delta_1$ 
will again mix in some phase-shift relations \cite{fu}. 
\end{itemize}
We verified all the above relations also with the explicit numerical evaluation of $Z^{\mathbf{d}}_{lm}$ using the expression in Appendix A: in particular $Z^{\mathbf{d}}_{10}(m_1\not = m_2)\not =0$ as already found in \cite{fu}, 
while  $Z^{\mathbf{d}}_{10}(m_1=m_2)=0$ as required by the symmetries of $D_{4h}$ for mass-degenerate case. 

\subsubsection{The matrix ${\cal M}_{lm,l'm'}^{\mathbf{d}}$ for $\mathbf{d}=e_z$}

The above relations for $Z_{lm}^{\mathbf{d}}$ or  $w_{lm}$ (\ref{w}) simplify the general ${\cal M}$ (\ref{M_gen}) for $\mathbf{d}=e_z$  to\footnote{We find an additional factor $\sqrt{3}$ in front of $w_{10}$ with respect to \cite{fu}.}
\begin{equation}
\label{M_001}
M={\cal M}_{lm,l'm'}^{\mathbf{d}} = \bordermatrix{~ & 00        & 10                    & 11                       & 1-1\cr
  00 & w_{00} & i\sqrt{3} w_{10} & 0  & 0 \cr
  10 & -i\sqrt{3}w_{10}      & w_{00}+2w_{20} & 0 & 0 \cr
  11 & 0 & 0 & w_{00}-w_{20} & 0 \cr
 1-1 & 0 & 0 & 0 & w_{00}-w_{20} \cr}
\end{equation}
where all $w_{l0}$ are real as indicated in the previous subsection. 

%%%%%%%%%%%%%%

\subsubsection{Phase shift relations for $\mathbf{d}=e_z$}

The phase shift relation are obtained from the determinant condition (\ref{det}) using the matrix $M$ (\ref{M_001}). 
This matrix is already in the block-diagonal form, that can not be reduced further. The determinant condition requires that determinant for each of the block is equal to zero. \\

 {\bf Extracting $P$-wave phase shift from irrep $E$}

\vspace{0.2cm}

The determinant condition for each of two $1\times 1$ blocks leads to $\tan(\delta_1)=1/M^{E}$ with $M^{E}=w_{00}-w_{20}$
\begin{equation}
\label{ph_001_E}
 \mathbf{d}=e_z~,\  E\ \mathrm{of}\ C_{4v} \ : \quad \tan\delta_1(p^*)=\frac{\pi^{3/2}~\gamma ~q}{Z_{00}^{\mathbf{d}}(1;q^2)-\tfrac{1}{\sqrt{5}}~q^{-2}~Z_{20}^{\mathbf{d}}(1;q^2)}   
 \end{equation}
 Basis vectors $Y_{11}$ and $Y_{1-1}$ form two-dimensional irreducible representation $E$ (Table \ref{tab_c4v_char}) 
 and interpolators that transform according to one of those two will naturally obey the same phase shift relation. 
 Note that the more conventional basis  vectors $x\propto Y_{11}-Y_{1-1}$ and $y\propto Y_{11}+Y_{1-1}$ will lead to the same matrix $M$ (\ref{M_001}) and therefore the same phase shift relation (\ref{ph_001_E}) applies for them. 
 Our interpolators in $E$ representation will transform like $x$ or $y$. 
 
 The phase shift relation for irrep $E$ (\ref{ph_001_E}) agree  with the expression in \cite{rho_pacscs} for the case $m_1=m_2$. \\

 {\bf Extracting  $S$-wave phase shift from irrep $A_1$}

\vspace{0.2cm}

The $2\times 2$ block of $M$ (\ref{M_001}) spans the  basis $Y_{00}\propto 1$ and $Y_{10}\propto z$, which are both invariant under all $\hat R\in C_{2v}$ and therefore belong to irrep $A_1$ (Table \ref{tab_c4v_char}). 
 The determinant condition (\ref{det}) for this $2\times 2$ block requires
\begin{align}
\label{ph_001_A1}
&\mathbf{d}=e_z~,  \  A_1\ \mathrm{of}\ C_{4v}\ :\\
&[e^{2i\delta_0(p^*)}(w_{00}-i)-(w_{00}+i)][e^{2i\delta_1(p^*)}(w_{00}+2w_{20}-i)-(w_{00}+2w_{20}+i)]\nonumber\\
 &=3|w_{10}|^2(e^{2i\delta_0(p^*)}-1)(e^{2i\delta_1(p^*)}-1)\nonumber
\end{align}
where $w_{ij}$ (\ref{w}) depend on $Z_{lm}^{\mathbf{d}}(1;q^2)$ and $\mathbf{d}=e_z$. 

If we know the energy $E$ of two particles in irrep $A_1$ on the lattice, the relation (\ref{ph_001_A1}) presents one relation with two unknowns $\delta_0(p^*)$ and $\delta_1(p^*)$, which was already noticed in \cite{fu}. A reliable extracting of $\delta_0(p^*)$ from (\ref{ph_001_A1}) 
\begin{equation}
\mathbf{d}=e_z~,  \  A_1\ \mathrm{of}\ C_{4v}\ :\quad \tan\delta_0(p^*)=\frac{1-\tan\delta_1(p^*)~[w_{00}+2w_{20}]}{w_{00}-\tan\delta_1(p^*)~[w_{00}^2+2w_{00}w_{20}- 3 w_{10}^2]}\nonumber
\end{equation}
is challenging since one needs the value of $\delta_1(p^*)$ at the same $p^*$. We proposed several strategies for estimating this in the corresponding section on $\mathbf{d}=e_x+e_y$  and the same strategies may be used also for $\mathbf{d}=e_z$. 
The only difference is that the 1D irreps  $B_{2,3}$ are replaced by the 2D irrep $E$. 

\subsubsection{Quark-antiquark and meson-meson interpolators for $\mathbf{d}=e_z$}

We list examples of the  quark-antiquark  and two-pseudoscalar   interpolators in the lattice frame (that transform according to $E$ or $A_1$ in CMF)  
\begin{align}
\label{interpolators_c4v}
({\cal O}^{\bar qq}_{E})_k&=V_k(e_z)~,\qquad\qquad\qquad k=x,\ y \\
({\cal O}^{\bar qq}_{A_1})^I&=V_z(e_z)\nonumber\\
({\cal O}^{\bar qq}_{A_1})^{II}&=S(e_z)\nonumber\\
~&\nonumber\\
({\cal O}^{P_1P_2}_{E})_k^{I} &=P_1(e_z+e_k)P_2(-e_k)-P_1(e_z-e_k)P_2(e_k)~,\qquad k=x,\ y\nonumber \\
({\cal O}^{P_1P_2}_{E})_k^{II} &=P_1(e_z+u_k)P_2(-u_k)-P_1(e_z-u_k)P_2(u_k)~,\qquad u_k=e_x+e_y,\ e_x-e_y\nonumber \\
~&\nonumber\\
({\cal O}^{P_1P_2}_{A_{1}})^{I} &=P_1(e_z)P_2(0)\nonumber\\
({\cal O}^{P_1P_2}_{A_1})^{II} &=P_1(e_z+e_x)P_2(-e_x)+P_1(e_z-e_x)P_2(e_x)+P_1(e_z+e_y)P_2(-e_y)+P_1(e_z-e_y)P_2(e_y)\nonumber\\
({\cal O}^{P_1P_2}_{A_1})^{III} &=P_1(e_z+e_x+e_y)P_2(-e_x-e_y)+P_1(e_z+e_x-e_y)P_2(-e_x+e_y)\nonumber\\
\qquad\qquad &+P_1(e_z-e_x+e_y)P_2(e_x-e_y)+P_1(e_z-e_x-e_y)P_2(e_x+e_y)\nonumber
\end{align}
and  the  analogous interpolators where the flavors $P_1$ and $P_2$ are interchanged. The representation $E$ is two-dimensional, so index $k$ in $({\cal O}_E)_k$ carries two values. The interpolators (\ref{interpolators_c4v}) are  expressed in terms of the currents (\ref{currents}) and  are constructed in analogous way as for $\mathbf{d}=e_x+e_y$. We list all the $P_1P_2$ interpolators where 
$p_{1}\leq \sqrt{3}~ \tfrac{2\pi}{L}$ and $p_{2}\leq \sqrt{3}~ \tfrac{2\pi}{L}$.

\vspace{0.2cm}
 
Let us consider the  transformation of (\ref{interpolators_c4v}) properties on the example of 
$({\cal O}^{P_1P_2}_{E})_k^I$, which becomes
$({\cal O}^{P_1P_2}_{E\ k})^I_{CMF} =P_1(ce_z+e_k)P_2(-ce_z-e_k)-P_1(ce_z-e_k)P_2(-ce_z+e_k)$   after the boost to CMF where $c=\gamma^{-1}(1-\tfrac{1}{2}A)$. 
The CMF interpolator transforms like $e_{k=x,y}$, as can be understood from the discussion for $\mathbf{d}=e_x+e_y$ case, so  the CMF interpolator transforms according to $2D$ irrep $E$ (see Table \ref{tab_c4v_char}). 

We note that the  interpolator ${\cal O}^{\bar qq}_{E}$ has been applied for the study for $\rho$ resonance \cite{rho_pacscs} where $m_1=m_2=m_\pi$, while ${\cal O}^{P_1P_2}_{E}$ has been written down\footnote{$I\in D_{4h}$ in the $m_1=m_2$ case, so ${\cal O}^{PP}_{E}$ is anti-symmetrized with respect to both particles in \cite{rho_pacscs}. }  but not employed. 

\section{Conclusions}

We derived  the relations that allow the lattice QCD extraction of the scattering phase shift $\delta_l(s)$ from two-particle energy $E$ in the finite box. We consider the  scattering of two particles with $m_1\not = m_2$  and with total momentum $\mathbf{P}\not = 0$.   The simulation of the system with  $\mathbf{P}\not =0$ will be important in practice as it will allow the extraction of $\delta_l(s)$ at several values of $s=E^2-\mathbf{P}^2$. 

We find that the $P$-wave phase shift can be extracted from the irreducible representation $E$ of $C_{4v}$ for $\mathbf{P}=(2\pi/L)e_z$ (\ref{ph_001_E}), or from the irreducible representations $B_{2,3}$ of $C_{2v}$ for $\mathbf{P}=(2\pi/L)(e_x+e_y)$ (\ref{ph_110_B3},\ref{ph_110_B2}). To be more specific, these relations allow a reliable extraction of $\delta_1(s)$ when  $s$ is below inelastic threshold, when  $\delta_{l\geq 2}(s)$ can be neglected and when $L$ is large enough that powers of $e^{-m_\pi L}$ can be neglected. If these conditions are not satisfied, one needs to generalize the  phase shift relations presented here. 

The reliable extraction of the $S$-wave phase shift from a simulation with  $\mathbf{P}\not =0$ will be challenging even if the above three conditions are fulfilled. 
The reason is that $\delta_0(s)$  appears together with $\delta_1(s)$ in the $A_1$ representation when $\mathbf{P}\not =0 $ and $m_1\not = m_2$. This mixing happens since the inversion is not the symmetry of the two-particle system in CMF. We propose several strategies that allow an estimate of $\delta_0(s)$ at $\mathbf{P}\not =0$ in spite of this problem. We expect that a more  reliable extraction of $S$-wave phase shift for two particles with $m_1\not =m_2$ needs to be performed using a simulation with $\mathbf{P}=0$ at several values of the lattice size $L$; in this case $\delta_0$ mixes only with $\delta_{l\geq 4}$ and these can be safely neglected. 
  
Besides the phase shift relations, we wrote down  also the quark-antiquark and meson-meson interpolators that transform according to the considered irreducible representations. These can be used in actual simulations. 

\vspace{1cm}
 
{\bf Acknowledgments}

\vspace{0.1cm}

 We would kindly like to thank  C.B. Lang for reading the manuscript. 
We would like to acknowledge valuable discussions with X. Feng, C.B. Lang, D. Mohler, A. Rusetsky and M. Savage.

%%%%%%%%%%%%%%%%%%%%%%%%%%%%%%%%%%%%%%%%%%%%%%%%%%%%%%%%%%%%%%%%%%%%%%
%%%%%%%%%%%%%%%%%%%%%%%%%%%%%%%%%%%%%%%%%%%%%%%%%%%%%%%%%%%%%%%%%%%%%%

\vspace{2cm}

\newpage

\appendix

\section*{Appendix A: Evaluation of $Z^{\mathbf{d}}_{lm}(1;q^2)$ for $m_1\not = m_2$}

In this appendix we derive a form of the  generalized function $Z_{lm}^{\mathbf{d}}(1;q^2)$  that is appropriate for numerical evaluation. 
We consider the most general case $m_1\not = m_2$,  $\mathbf{d}=\tfrac{2\pi}{L}\mathbf{P} \not = 0$ and general $l$ and $m$, which has not been 
considered before. Some parts of our derivation  are similar to  Appendix A of \cite{yamazaki}, done for $l=m=0$ and $m_1=m_2$, and to  Appendix B of \cite{fu}, done for  $l=1$ and $m=0$. 

The $Z_{lm}^{\mathbf{d}}(s;q^2)$ for the general case of $m_1\not = m_2$ and $\mathbf{d}=\tfrac{2\pi}{L}\mathbf{P} \not = 0$ is defined in (\ref{Z_def})  
\begin{equation}
\label{Zlm_def}
Z_{lm}^{\mathbf{d}}(s;q^2)\equiv \sum_{\mathbf{r}\in P_{d}} \frac{{\cal Y}_{lm}(\mathbf{r})}{(|\mathbf{r}|^2-q^2)^s}\ , \qquad q=\frac{L}{2\pi}~ p^*~,\quad {\cal Y}_{lm}(\mathbf{r})\equiv r^l Y_{lm}(\theta,\phi) \ ,
\end{equation}
where the relations for the phase shift depend on $Z_{lm}^{\mathbf{d}}(1;q^2)$ evaluated at $s=1$. Here $q^2$ is real and can be positive or negative (\ref{helmholtz2}). The sum goes over the mesh $P_{d}$ defined in (\ref{P_d}) and plotted in Fig. \ref{fig_mesh_110}c for $\mathbf{d}=e_x+e_y$ and in  \ref{fig_mesh_001}c for $\mathbf{d}=e_z$.

The sum is finite at $s=1$ for every $l$ and $m$ except for $l=m=0$, and we will derive the expression that converges faster than (\ref{Zlm_def}), and is appropriate for numerical evaluation. We will show that  sum converges only for $s>3/2$ (but not $s=1$) in case of $l=m=0$. 
The divergence that appears  for $s=1$  will be  exactly equal to the divergence that appears in the infinite volume. Since the phase-shift relations depend on the finite volume shift with respect to the infinite volume, we will get rid of the divergence by the analytic continuation from $s>3/2$ to $s=1$. 

First we express $1/(r^2-q^2)^s$ using the definition of the Gamma function $\int_0^\infty dt~ t^{s-1}~e^{-ta}=\Gamma(s)/a^s$ and then split the integral to two parts
\begin{align}
\label{Zlm_1}
Z_{lm}^{\mathbf{d}}(s;q^2)&= \frac{1}{\Gamma(s)}\sum_{\mathbf{r}\in P_{\mathbf{d}}} {\cal Y}_{lm}(\mathbf{r}) \int _0^\infty dt~t^{s-1} e^{-t(r^2-q^2)} \nonumber \\
&=\frac{1}{\Gamma(s)}\sum_{\mathbf{r}\in P_{d}} {\cal Y}_{lm}(\mathbf{r})~\bigl\{~\int _0^1 dt~t^{s-1} e^{-t(r^2-q^2)}~+~\int _1^\infty dt~t^{s-1} e^{-t(r^2-q^2)}~\bigr\}\ .
\end{align}
The integral in the second term is finite at $s=1$, it is easily evaluated, and renders faster convergence than the original sum 
\begin{equation}
\label{second_term}
\mathrm{second\ term}=\sum_{\mathbf{r}\in P_{d}} {\cal Y}_{lm}(\mathbf{r}) \frac{1}{\Gamma(s)} \int _1^\infty dt~t^{s-1} e^{-t(r^2-q^2)} \ \stackrel {s=1}{ \longrightarrow} ~\sum_{\mathbf{r}\in P_{d}} {\cal Y}_{lm}(\mathbf{r}) \frac{e^{-(r^2-q^2)}}{r^2-q^2}\ .
\end{equation}

The first term (\ref{Zlm_1}) contains the sum 
   $\sum_{\mathbf{r} \in P_d}F(\mathbf{r})$, which   is equivalent to the sum $\sum_{\mathbf{n}\in Z^3} F(\mathbf{r}(\mathbf{n}))$ and we express  it  using the Poisson summation formula  
\begin{equation}
\label{poisson}
\sum_{\mathbf{n}\in Z^3} f(\mathbf{n})=\sum_{\mathbf{n}\in Z^3} \int d^3x~f(\mathbf{x})~e^{i2\pi \mathbf{n}\cdot \mathbf{x}}
\end{equation}
leading to
\begin{equation}
\label{first_term}
\mathrm{first\ term}=\frac{1}{\Gamma(s)} \int _0^1 dt~t^{s-1} e^{tq^2} \sum_{\mathbf{n}\in Z^3} f_{\mathbf{n}}~,\quad f_{\mathbf{n}}\equiv \int d^3x~{\cal Y}_{lm}(\mathbf{r}(\mathbf{x}))~ e^{-t|\mathbf{r}(\mathbf{x})|^2~+~i2\pi \mathbf{n}\cdot \mathbf{x}} \ 
\end{equation}
with $\mathbf{r}(\mathbf{x})=\hat \gamma^{-1}(\mathbf{x}-\tfrac{1}{2}A\mathbf{d})$ (\ref{P_d}). 
 We change the integration variable  from $\mathbf{x}$ to $\mathbf{r}$ using $d^3x=det(J)d^3r=\gamma d^3r$ and separate terms that depend only on $r$  using ${\cal Y}_{lm}(\mathbf{r})= r^l Y_{lm}(\theta,\phi)$.  Applying  $\mathbf{x}=\hat \gamma \mathbf{r}+\tfrac{1}{2}A\mathbf{d}$  (\ref{P_d}) the term dependent on $A$ factorizes 
\begin{equation}
\label{f}
f_{\mathbf{n}}\equiv \gamma~ e^{i\pi A \mathbf{n\cdot d}}  \int_0^\infty r^2 dr ~ e^{-tr^2} ~r^l~\int_0^\pi \sin\theta d\theta\int_0^{2\pi} d\phi~ Y_{lm}(\theta,\phi)~e^{-i\mathbf{k}\cdot \mathbf{r}}   
\end{equation}
with $\mathbf{k}\equiv -2\pi \hat \gamma^T \mathbf{n}$. We insert the well known relation for $e^{-i\mathbf{k}\cdot \mathbf{r}}$

\begin{equation}
e^{-i\mathbf{k}\cdot \mathbf{r}}=4\pi\sum_{l'=0}^\infty \sum_{m'=-l'}^{l'}  (-i)^{l'} ~Y_{l'm'}(\theta_k,\phi_k)~Y_{l'm'}(\theta,\phi)^*~j_{l'}(kr)~.
\end{equation}
The integral $\int_0^\pi \sin\theta d\theta~\int_0^{2\pi} d\phi Y_{lm}(\theta,\phi)Y^*_{l'm'}(\theta,\phi)=\delta_{ll'}\delta_{mm'}$ simplifies (\ref{f})   to 
\begin{align}
f_{\mathbf{n}}= \gamma ~4\pi~(-i)^l~ (-1)^{A\mathbf{n\cdot d}} ~ Y_{lm}(\theta_k,\phi_k) \int_0^\infty ~dr ~ r^2~ e^{-tr^2} ~r^l~ j_l(kr)~.
\end{align}
The remaining integral can be evaluated with Mathematica
\begin{align}
f_{\mathbf{n}}= \gamma (-i)^l~ (-1)^{A\mathbf{n\cdot d}} ~ \biggl(\frac{k}{2t}\biggr)^l~Y_{lm}(\theta_k,\phi_k) \biggl(\frac{\pi}{t}\biggr)^{3/2}  e^{-k^2/4t}~
\end{align} 
and we apply $(k/2t)^l~Y_{lm}(\theta_k,\phi_k)={\cal Y}_{lm}(\mathbf{k}/2t)={\cal Y}_{lm}(-\pi\hat \gamma\mathbf{n}/t)$. Inserting this $f_{\mathbf{n}}$ to (\ref{first_term}) we get
\begin{equation}
\mathrm{first\ term}=\frac{1}{\Gamma(s)} \int _0^1 dt~t^{s-1} e^{tq^2} \sum_{\mathbf{n}\in Z^3} \gamma (-i)^l~ (-1)^{A\mathbf{n\cdot d}} ~ {\cal Y}_{lm}(-\frac{\pi\hat \gamma \mathbf{n}}{t}) \biggl(\frac{\pi}{t}\biggr)^{3/2}  e^{-(\pi \hat \gamma \mathbf{n})^2/t}~.
\end{equation}
In the case of $s=1$, this integral over $t$ is finite for all $\mathbf{n}$ except for $\mathbf{n}=0$. The $\mathbf{n}=0$  divergence occurs only for $l=m=0$ since ${\cal Y}_{lm}(\mathbf{n}=0)\propto \delta_{l0}\delta_{m0}$. The term with $\mathbf{n}=0$  is the infinite volume  $f_{\mathbf{n}=0}=\int d^3x f(\mathbf{x})$  analog of  $\sum_{\mathbf{n}} f(\mathbf{n})$ in the Poisson's formula (\ref{poisson}) and is finite only for $s>3/2$. In order to get rid of the divergence, that cancels in the difference between the finite and infinite volume result anyway, we split the $\mathbf{n}=0$ term in two parts
\begin{equation}
\frac{1}{\Gamma(s)} \int _0^1 dt~t^{s-5/2} e^{tq^2} = \frac{1}{\Gamma(s)} \biggl[\int _0^1 dt~t^{s-5/2} (e^{tq^2}-1)+\int _0^1 dt~t^{s-5/2}\biggr]~.
\end{equation}
The first integral is finite for $s=1$, while the second integral $\int_0^1  t^{s-5/2}dt\stackrel{s>3/2}{ =}\tfrac{1}{s-3/2}\stackrel{s\to  1}{ \longrightarrow}-2 $ is finite only for $s>3/2$, but we analytically continue it to $s=1$.

Collecting (\ref{second_term}) as well as convergent and divergent piece of (\ref{first_term}) to get (\ref{Zlm_1}), we get  finally 
\begin{align}
Z_{lm}^{\mathbf{d}}(1;q^2)&=  \gamma \int _0^1 dt~ e^{tq^2} \sum_{\mathbf{n}\in Z^3,\mathbf{n}\not=0}   (-1)^{A\mathbf{n\cdot d}}~(-i)^l~ {\cal Y}_{lm}(-\frac{\pi\hat \gamma \mathbf{n}}{t}) (\frac{\pi}{t})^{3/2}  e^{-(\pi \hat \gamma \mathbf{n})^2/t}\nonumber \\
 &+\gamma  \int_0^1 dt~(e^{tq^2}-1)\biggl(\frac{\pi}{t}\biggr)^{3/2} \frac{1}{\sqrt{4\pi}}\delta_{l0}\delta_{m0}-\gamma\pi\delta_{l0}\delta_{m0}\nonumber\\
&+\sum_{\mathbf{r}\in P_{d}} {\cal Y}_{lm}(\mathbf{r}) \frac{e^{-(r^2-q^2)}}{r^2-q^2}\ 
\end{align}
which is used for our numerical evaluation and converges rapidly for $l,m,\mathbf{d}$  of our interest. It is applicable for $q^2>0$ and $q^2<0$. We verified numerically that this $Z_{lm}^{\mathbf{d}}$ respects all the relations listed in the main text, that follow from discrete symmetries at $\mathbf{d}=e_x+e_y$ or $\mathbf{d}=e_z$.   
 
% we could comment on exchanging m1<->m2

In the special case $m_1=m_2$, our result agrees with the result in \cite{feng_proc}, which was presented for $m_1=m_2$ without derivation\footnote{Note that 
$Z_{lm}^{\mathbf{d}}$ in \cite{feng_proc} is defined to be complex conjugate of ours.}. 
We also verified that such $Z^{\mathbf{d}}_{lm}$  numerically agrees with  $Z^{\mathbf{d}}_{lm}$ obtained for $m_1=m_2$ via $c_{lm}$ as proposed by \cite{sharpe}.

\end{document}